\documentclass[reprint,superscriptaddress,twocolumn,showpacs,preprintnumbers,amsmath,amssymb]{revtex4-1}

\usepackage{multirow}
\usepackage{graphicx}
\usepackage{epsfig}

\usepackage[mathlines]{lineno}

\begin{document}

\def \nobreakseq {\nobreak \hskip 0pt \hbox}


\title{Measurement of the $2\nu\beta\beta$ Decay Half-Life and Search for the $0\nu\beta\beta$ Decay of $^{116}$Cd with the NEMO-3 Detector}

\author{R.~Arnold}
\affiliation{IPHC, ULP, CNRS/IN2P3\nobreakseq{,} F-67037 Strasbourg, France}
\author{C.~Augier}
\affiliation{LAL, Universit\'e Paris-Sud\nobreakseq{,} CNRS/IN2P3\nobreakseq{,} F-91405 Orsay\nobreakseq{,} France}
\author{J.D.~Baker}\thanks{Deceased}
\affiliation{Idaho National Laboratory\nobreakseq{,} Idaho Falls, ID 83415, U.S.A.}
\author{A.S.~Barabash}
\affiliation{NRC "Kurchatov Institute", ITEP, 117218 Moscow, Russia}
\author{A.~Basharina-Freshville}
\affiliation{UCL, London WC1E 6BT\nobreakseq{,} United Kingdom}
\author{S.~Blondel}
\affiliation{LAL, Universit\'e Paris-Sud\nobreakseq{,} CNRS/IN2P3\nobreakseq{,} F-91405 Orsay\nobreakseq{,} France}
\author{S.~Blot}
\affiliation{University of Manchester\nobreakseq{,} Manchester M13 9PL\nobreakseq{,}~United Kingdom}
\author{M.~Bongrand}
\affiliation{LAL, Universit\'e Paris-Sud\nobreakseq{,} CNRS/IN2P3\nobreakseq{,} F-91405 Orsay\nobreakseq{,} France}
\author{D.~Boursette}
\affiliation{LAL, Universit\'e Paris-Sud\nobreakseq{,} CNRS/IN2P3\nobreakseq{,} F-91405 Orsay\nobreakseq{,} France}
\author{V.~Brudanin}
\affiliation{JINR, 141980 Dubna, Russia}
\affiliation{National Research Nuclear University MEPhI, 115409 Moscow, Russia}
\author{J.~Busto}
\affiliation{Aix Marseille Universit\'e, CNRS, CPPM, Marseille, France}
\author{A.J.~Caffrey}
\affiliation{Idaho National Laboratory\nobreakseq{,} Idaho Falls, ID 83415, U.S.A.}
\author{S.~Calvez}
\affiliation{LAL, Universit\'e Paris-Sud\nobreakseq{,} CNRS/IN2P3\nobreakseq{,} F-91405 Orsay\nobreakseq{,} France}
\author{M.~Cascella} 
\affiliation{UCL, London WC1E 6BT\nobreakseq{,} United Kingdom}
\author{C.~Cerna}
\affiliation{CENBG\nobreakseq{,} Universit\'e de Bordeaux\nobreakseq{,} CNRS/IN2P3\nobreakseq{,} F-33175 Gradignan\nobreakseq{,} France}
\author{J.~P.~Cesar}
\affiliation{University of Texas at Austin\nobreakseq{,} Austin\nobreakseq{,} TX 78712\nobreakseq{,}~U.S.A.}
\author{A.~Chapon}
\affiliation{LPC Caen\nobreakseq{,} ENSICAEN\nobreakseq{,} Universit\'e de Caen\nobreakseq{,} CNRS/IN2P3\nobreakseq{,} F-14050 Caen\nobreakseq{,} France}
\author{E.~Chauveau}
\affiliation{University of Manchester\nobreakseq{,} Manchester M13 9PL\nobreakseq{,}~United Kingdom}
\author{A.~Chopra} 
\affiliation{UCL, London WC1E 6BT\nobreakseq{,} United Kingdom}
\author{D.~Duchesneau}
\affiliation{LAPP, Universit\'e de Savoie Mont Blanc\nobreakseq{,} CNRS/IN2P3\nobreakseq{,} F-74941 Annecy-le-Vieux\nobreakseq{,} France}
\author{D.~Durand}
\affiliation{LPC Caen\nobreakseq{,} ENSICAEN\nobreakseq{,} Universit\'e de Caen\nobreakseq{,} CNRS/IN2P3\nobreakseq{,} F-14050 Caen\nobreakseq{,} France}
\author{V.~Egorov}
\affiliation{JINR, 141980 Dubna, Russia}
\author{G.~Eurin}
\affiliation{LAL, Universit\'e Paris-Sud\nobreakseq{,} CNRS/IN2P3\nobreakseq{,} F-91405 Orsay\nobreakseq{,} France}
\affiliation{UCL, London WC1E 6BT\nobreakseq{,} United Kingdom}
\author{J.J.~Evans}
\affiliation{University of Manchester\nobreakseq{,} Manchester M13 9PL\nobreakseq{,}~United Kingdom}
\author{L.~Fajt}
\affiliation{Institute of Experimental and Applied Physics\nobreakseq{,} Czech Technical University in Prague\nobreakseq{,} CZ-12800 Prague\nobreakseq{,} Czech Republic}
\author{D.~Filosofov}
\affiliation{JINR, 141980 Dubna, Russia}
\author{R.~Flack}
\affiliation{UCL, London WC1E 6BT\nobreakseq{,} United Kingdom}
\author{X.~Garrido}
\affiliation{LAL, Universit\'e Paris-Sud\nobreakseq{,} CNRS/IN2P3\nobreakseq{,} F-91405 Orsay\nobreakseq{,} France}
\author{H.~G\'omez}
\affiliation{LAL, Universit\'e Paris-Sud\nobreakseq{,} CNRS/IN2P3\nobreakseq{,} F-91405 Orsay\nobreakseq{,} France}
\author{B.~Guillon}
\affiliation{LPC Caen\nobreakseq{,} ENSICAEN\nobreakseq{,} Universit\'e de Caen\nobreakseq{,} CNRS/IN2P3\nobreakseq{,} F-14050 Caen\nobreakseq{,} France}
\author{P.~Guzowski}
\affiliation{University of Manchester\nobreakseq{,} Manchester M13 9PL\nobreakseq{,}~United Kingdom}
\author{R.~Hod\'{a}k}
\affiliation{Institute of Experimental and Applied Physics\nobreakseq{,} Czech Technical University in Prague\nobreakseq{,} CZ-12800 Prague\nobreakseq{,} Czech Republic}
\author{A.~Huber}
\affiliation{CENBG\nobreakseq{,} Universit\'e de Bordeaux\nobreakseq{,} CNRS/IN2P3\nobreakseq{,} F-33175 Gradignan\nobreakseq{,} France}
\author{P.~Hubert}
\affiliation{CENBG\nobreakseq{,} Universit\'e de Bordeaux\nobreakseq{,} CNRS/IN2P3\nobreakseq{,} F-33175 Gradignan\nobreakseq{,} France}
\author{C.~Hugon}
\affiliation{CENBG\nobreakseq{,} Universit\'e de Bordeaux\nobreakseq{,} CNRS/IN2P3\nobreakseq{,} F-33175 Gradignan\nobreakseq{,} France}
\author{S.~Jullian}
\affiliation{LAL, Universit\'e Paris-Sud\nobreakseq{,} CNRS/IN2P3\nobreakseq{,} F-91405 Orsay\nobreakseq{,} France}
\author{A.~Klimenko}
\affiliation{JINR, 141980 Dubna, Russia}
\author{O.~Kochetov}
\affiliation{JINR, 141980 Dubna, Russia}
\author{S.I.~Konovalov}
\affiliation{NRC "Kurchatov Institute", ITEP, 117218 Moscow, Russia}
\author{V.~Kovalenko}
\affiliation{JINR, 141980 Dubna, Russia}
\author{D.~Lalanne}
\affiliation{LAL, Universit\'e Paris-Sud\nobreakseq{,} CNRS/IN2P3\nobreakseq{,} F-91405 Orsay\nobreakseq{,} France}
\author{K.~Lang}
\affiliation{University of Texas at Austin\nobreakseq{,} Austin\nobreakseq{,} TX 78712\nobreakseq{,}~U.S.A.}
\author{Y.~Lemi\`ere}
\affiliation{LPC Caen\nobreakseq{,} ENSICAEN\nobreakseq{,} Universit\'e de Caen\nobreakseq{,} CNRS/IN2P3\nobreakseq{,} F-14050 Caen\nobreakseq{,} France}
\author{T.~Le~Noblet}
\affiliation{LAPP, Universit\'e de Savoie Mont Blanc\nobreakseq{,} CNRS/IN2P3\nobreakseq{,} F-74941 Annecy-le-Vieux\nobreakseq{,} France}
\author{Z.~Liptak}
\affiliation{University of Texas at Austin\nobreakseq{,} Austin\nobreakseq{,} TX 78712\nobreakseq{,}~U.S.A.}
\author{X.~R.~Liu} 
\affiliation{UCL, London WC1E 6BT\nobreakseq{,} United Kingdom}  
\author{P.~Loaiza}
\affiliation{LAL, Universit\'e Paris-Sud\nobreakseq{,} CNRS/IN2P3\nobreakseq{,} F-91405 Orsay\nobreakseq{,} France}
\author{G.~Lutter}
\affiliation{CENBG\nobreakseq{,} Universit\'e de Bordeaux\nobreakseq{,} CNRS/IN2P3\nobreakseq{,} F-33175 Gradignan\nobreakseq{,} France}
\author{M.~Macko}
\affiliation{FMFI,~Comenius~University \nobreakseq{,}~SK-842~48~Bratislava\nobreakseq{,}~Slovakia}
\affiliation{CENBG\nobreakseq{,} Universit\'e de Bordeaux\nobreakseq{,} CNRS/IN2P3\nobreakseq{,} F-33175 Gradignan\nobreakseq{,} France}
\author{C.~Macolino}
\affiliation{LAL, Universit\'e Paris-Sud\nobreakseq{,} CNRS/IN2P3\nobreakseq{,} F-91405 Orsay\nobreakseq{,} France}
\author{F.~Mamedov}
\affiliation{Institute of Experimental and Applied Physics\nobreakseq{,} Czech Technical University in Prague\nobreakseq{,} CZ-12800 Prague\nobreakseq{,} Czech Republic}
\author{C.~Marquet}
\affiliation{CENBG\nobreakseq{,} Universit\'e de Bordeaux\nobreakseq{,} CNRS/IN2P3\nobreakseq{,} F-33175 Gradignan\nobreakseq{,} France}
\author{F.~Mauger}
\affiliation{LPC Caen\nobreakseq{,} ENSICAEN\nobreakseq{,} Universit\'e de Caen\nobreakseq{,} CNRS/IN2P3\nobreakseq{,} F-14050 Caen\nobreakseq{,} France}
\author{B.~Morgan}
\affiliation{University of Warwick\nobreakseq{,} Coventry CV4 7AL\nobreakseq{,} United Kingdom}
\author{J.~Mott}
\affiliation{UCL, London WC1E 6BT\nobreakseq{,} United Kingdom}
\author{I.~Nemchenok}
\affiliation{JINR, 141980 Dubna, Russia}
\author{M.~Nomachi}
\affiliation{Osaka University\nobreakseq{,} 1-1 Machikaney arna Toyonaka\nobreakseq{,} Osaka 560-0043\nobreakseq{,} Japan}
\author{F.~Nova}
\affiliation{University of Texas at Austin\nobreakseq{,} Austin\nobreakseq{,} TX 78712\nobreakseq{,}~U.S.A.}
\author{F.~Nowacki}
\affiliation{IPHC, ULP, CNRS/IN2P3\nobreakseq{,} F-67037 Strasbourg, France}
\author{H.~Ohsumi}
\affiliation{Saga University\nobreakseq{,} Saga 840-8502\nobreakseq{,} Japan}
\author{R.B.~Pahlka}
\affiliation{University of Texas at Austin\nobreakseq{,} Austin\nobreakseq{,} TX 78712\nobreakseq{,}~U.S.A.}
\author{F.~Perrot}
\affiliation{CENBG\nobreakseq{,} Universit\'e de Bordeaux\nobreakseq{,} CNRS/IN2P3\nobreakseq{,} F-33175 Gradignan\nobreakseq{,} France}
\author{F.~Piquemal}
\affiliation{CENBG\nobreakseq{,} Universit\'e de Bordeaux\nobreakseq{,} CNRS/IN2P3\nobreakseq{,} F-33175 Gradignan\nobreakseq{,} France}
\affiliation{Laboratoire Souterrain de Modane\nobreakseq{,} F-73500 Modane\nobreakseq{,} France}
\author{P.~Povinec}
\affiliation{FMFI,~Comenius~University \nobreakseq{,}~SK-842~48~Bratislava\nobreakseq{,}~Slovakia}
\author{P.~P\v{r}idal}
\affiliation{Institute of Experimental and Applied Physics\nobreakseq{,} Czech Technical University in Prague\nobreakseq{,} CZ-12800 Prague\nobreakseq{,} Czech Republic}
\author{Y.A.~Ramachers}
\affiliation{University of Warwick\nobreakseq{,} Coventry CV4 7AL\nobreakseq{,} United Kingdom}
\author{A.~Remoto}
\affiliation{LAPP, Universit\'e de Savoie Mont Blanc\nobreakseq{,} CNRS/IN2P3\nobreakseq{,} F-74941 Annecy-le-Vieux\nobreakseq{,} France}
\author{J.L.~Reyss}
\affiliation{LSCE\nobreakseq{,} CNRS\nobreakseq{,} F-91190 Gif-sur-Yvette\nobreakseq{,} France}
\author{B.~Richards}
\affiliation{UCL, London WC1E 6BT\nobreakseq{,} United Kingdom}
\author{C.L.~Riddle}
\affiliation{Idaho National Laboratory\nobreakseq{,} Idaho Falls, ID 83415, U.S.A.}
\author{E.~Rukhadze}
\affiliation{Institute of Experimental and Applied Physics\nobreakseq{,} Czech Technical University in Prague\nobreakseq{,} CZ-12800 Prague\nobreakseq{,} Czech Republic}
\author{R.~Saakyan}
\affiliation{UCL, London WC1E 6BT\nobreakseq{,} United Kingdom}
\author{R.~Salazar}
\affiliation{University of Texas at Austin\nobreakseq{,} Austin\nobreakseq{,} TX 78712\nobreakseq{,}~U.S.A.}
\author{X.~Sarazin}
\affiliation{LAL, Universit\'e Paris-Sud\nobreakseq{,} CNRS/IN2P3\nobreakseq{,} F-91405 Orsay\nobreakseq{,} France}
\author{Yu.~Shitov}
\affiliation{JINR, 141980 Dubna, Russia}
\affiliation{Imperial College London\nobreakseq{,} London SW7 2AZ\nobreakseq{,} United Kingdom}
\author{L.~Simard}
\affiliation{LAL, Universit\'e Paris-Sud\nobreakseq{,} CNRS/IN2P3\nobreakseq{,} F-91405 Orsay\nobreakseq{,} France}
\affiliation{Institut Universitaire de France\nobreakseq{,} F-75005 Paris\nobreakseq{,} France}
\author{F.~\v{S}imkovic}
\affiliation{FMFI,~Comenius~University \nobreakseq{,}~SK-842~48~Bratislava\nobreakseq{,}~Slovakia}
\author{A.~Smetana}
\affiliation{Institute of Experimental and Applied Physics\nobreakseq{,} Czech Technical University in Prague\nobreakseq{,} CZ-12800 Prague\nobreakseq{,} Czech Republic}
\author{K.~Smolek}
\affiliation{Institute of Experimental and Applied Physics\nobreakseq{,} Czech Technical University in Prague\nobreakseq{,} CZ-12800 Prague\nobreakseq{,} Czech Republic}
\author{A.~Smolnikov}
\affiliation{JINR, 141980 Dubna, Russia}
\author{S.~S\"oldner-Rembold}
\affiliation{University of Manchester\nobreakseq{,} Manchester M13 9PL\nobreakseq{,}~United Kingdom}
\author{B.~Soul\'e}
\affiliation{CENBG\nobreakseq{,} Universit\'e de Bordeaux\nobreakseq{,} CNRS/IN2P3\nobreakseq{,} F-33175 Gradignan\nobreakseq{,} France}
\author{I.~\v{S}tekl}
\affiliation{Institute of Experimental and Applied Physics\nobreakseq{,} Czech Technical University in Prague\nobreakseq{,} CZ-12800 Prague\nobreakseq{,} Czech Republic}
\author{J.~Suhonen}
\affiliation{Jyv\"askyl\"a University\nobreakseq{,} FIN-40351 Jyv\"askyl\"a\nobreakseq{,} Finland}
\author{C.S.~Sutton}
\affiliation{MHC\nobreakseq{,} South Hadley\nobreakseq{,} Massachusetts 01075\nobreakseq{,} U.S.A.}
\author{G.~Szklarz}
\affiliation{LAL, Universit\'e Paris-Sud\nobreakseq{,} CNRS/IN2P3\nobreakseq{,} F-91405 Orsay\nobreakseq{,} France}
\author{J.~Thomas}
\affiliation{UCL, London WC1E 6BT\nobreakseq{,} United Kingdom}
\author{V.~Timkin}
\affiliation{JINR, 141980 Dubna, Russia}
\author{S.~Torre}
\affiliation{UCL, London WC1E 6BT\nobreakseq{,} United Kingdom}
\author{Vl.I.~Tretyak}
\affiliation{Institute for Nuclear Research\nobreakseq{,} MSP 03680 Kyiv\nobreakseq{,} Ukraine}
\author{V.I.~Tretyak}
\affiliation{JINR, 141980 Dubna, Russia}
\author{V.I.~Umatov}
\affiliation{NRC "Kurchatov Institute", ITEP, 117218 Moscow, Russia}
\author{I.~Vanushin}
\affiliation{NRC "Kurchatov Institute", ITEP, 117218 Moscow, Russia}
\author{C.~Vilela}
\affiliation{UCL, London WC1E 6BT\nobreakseq{,} United Kingdom}
\author{V.~Vorobel}
\affiliation{Charles University in Prague\nobreakseq{,} Faculty of Mathematics and Physics\nobreakseq{,} CZ-12116 Prague\nobreakseq{,} Czech Republic}
\author{D.~Waters}
\affiliation{UCL, London WC1E 6BT\nobreakseq{,} United Kingdom}
\author{A.~\v{Z}ukauskas}
\affiliation{Charles University in Prague\nobreakseq{,} Faculty of Mathematics and Physics\nobreakseq{,} CZ-12116 Prague\nobreakseq{,} Czech Republic}
\collaboration{NEMO-3 Collaboration}
\noaffiliation

\date{\today}

\begin{abstract}
The NEMO-3 experiment measured the half-life of the $2\nu\beta\beta$ decay and searched for the $0\nu\beta\beta$ decay of $^{116}$Cd.
Using 410~g of $^{116}$Cd installed in the detector with an exposure of 5.26~y, ($4968\pm74$)~events corresponding to the $2\nu\beta\beta$ decay of $^{116}$Cd to the ground state of $^{116}$Sn have been observed with a signal to background ratio of about 12.
The half-life of the $2\nu\beta\beta$ decay has been measured to be $ T_{1/2}^{2\nu}=[2.74\pm0.04\mbox{(stat.)}\pm0.18\mbox{(syst.)}]\times10^{19}$~y.
No events have been observed above the expected background while searching for $0\nu\beta\beta$ decay. 
The corresponding limit on the half-life is determined to be  $T_{1/2}^{0\nu} \ge 1.0 \times 10^{23}$~y at the 90\% C.L. which corresponds to an upper limit on the effective Majorana neutrino mass of $\langle m_{\nu} \rangle \le 1.4-2.5$~eV depending on the nuclear matrix elements considered.
Limits on other mechanisms generating $0\nu\beta\beta$ decay such as the exchange of R-parity violating supersymmetric particles, right-handed currents and majoron emission are also obtained. 
\end{abstract}
%
%
\maketitle
%
%
%
\section{Introduction \label{sec:intro}}
%
Experimental searches for neutrinoless double-beta decay ($0\nu\beta\beta$) are one of the most active research topics in neutrino physics. 
The observation of such a process is of major importance since it will prove the Majorana nature of neutrinos and may give access to their absolute mass scale. 
The Majorana nature of the neutrino would also have interesting implications in many extensions of the Standard Model of particle physics. 
For instance the see-saw mechanism requires the existence of a Majorana neutrino to explain the lightness of neutrino masses~\cite{Minkowski1977421,Yanagida:1980xy,GellMann1979,Mohapatra:1979ia}. 
A Majorana neutrino would also provide a natural framework for lepton number violation, and particularly for the leptogenesis process which may explain the observed matter-antimatter asymmetry of the universe~\cite{Fukugita198645}. 
The standard underlying mechanism behind neutrinoless double-beta decay is the exchange of a light Majorana neutrino. 
In this case, the decay rate can be written as:
\begin{equation}\label{eq:mm}
T^{0\nu}_{1/2} (A,Z)^{-1} = g_{A}^{4} G^{0\nu} (Q_{\beta\beta},Z) |M^{0\nu}(A,Z)|^{2} \left|\frac{ \langle m_{\nu} \rangle }{m_{e}} \right|^{2}
\end{equation}
where $g_{A}$ is the axial vector coupling constant, $G_{0\nu} $ is the kinematical phase space factor that depends on $Z$ and the nuclear
transition energy $Q_{\beta\beta}$, $M_{0\nu}$ is the Nuclear Matrix Element (NME), $m_{e}$ is the electron mass and $\langle m_{\nu} \rangle$ is the effective neutrino mass.
Other mechanisms could be involved in this process such as the exchange of supersymmetric particles via an R-parity violating coupling, the existence of right-handed currents in the electroweak interaction, or the emission of scalar bosons such as majorons~\cite{PhysRevD.76.093009,BAMERT199525,CARONE199385,Mohapatra2000143}. They would result in different energy and angular distributions of the emitted $\beta$ particles.
For a given mechanism and isotope, the $0\nu\beta\beta$ decay half-life depends on the phase space factor and on the NME. 
%
%
%
Due to different values of NMEs and phase space factors for different nuclei, the decay half-lives of different isotopes can differ by a few orders of magnitude.
%
%
%
\\ 
$^{116}$Cd is one of the best candidates to search for $0\nu\beta\beta$ decay given that the high $Q$-value of ($2813.50\pm0.13$)~keV~\cite{Rahaman2011412}  lies above most natural radioactive backgrounds.
The first observation of $2\nu\beta\beta$ decay in $^{116}$Cd was reported in 1995 by three independent experiments.
ELEGANT-V observed the two neutrino double-beta decay of $^{116}$Cd using natural and enriched cadmium foils sandwiched between drift chambers for trajectory reconstruction, plastic scintillators for energy measurement and sodium iodide scintillators to enhance background rejection. The half-life of $T^{2\nu}_{1/2} = 2.6^{+0.9}_{-0.5}\times10^{19}$~y was measured for the $2\nu\beta\beta$ decay and an upper limit of $T^{0\nu}_{1/2} > 2.9\times10^{21}$~y at $90$~\% C.L. was set for the $0\nu\beta\beta$ decay with an exposure of 0.02~kg$\times$y in terms of $^{116}$Cd mass~\cite{Kume}.
A compatible measurement was obtained with CdWO$_{4}$ crystals enriched to 83~\% in $^{116}$Cd at the Solotvina Underground laboratory. The half-life for the $2\nu\beta\beta$ decay was measured to be $T^{2\nu}_{1/2} = [2.7^{+0.5}_{-0.4}(\mbox{stat.})^{+0.9}_{-0.6}(\mbox{syst.})]\times10^{19}$~y and the upper limit on the $0\nu\beta\beta$ half-life was set to $T^{0\nu}_{1/2} > 2.9\times10^{22}$~y at 90~\% C.L. with an exposure of 0.13~kg$\times$y~\cite{Danevich199572}.
In the same year, the NEMO-2 experiment reported a measurement of the $2\nu\beta\beta$ half-life of $^{116}$Cd employing a tracking chamber and a plastic scintillator calorimeter~\cite{Arnold:1995jz}. The final measurement from NEMO-2 was performed in 1996 and reported a $2\nu\beta\beta$ half-life of $T^{2\nu}_{1/2} = [3.75\pm0.35(\mbox{stat.})\pm0.21 (\mbox{syst.})]\times10^{19}$~y~\cite{Arnold:1996wp} (later corrected to $T^{2\nu}_{1/2} = [2.9\pm0.3(\mbox{stat.})\pm0.2 (\mbox{syst.})]\times10^{19}$~y~\cite{PhysRevC.81.035501}) with an exposure of $0.11$~kg$\times$y. The lower limit on the  $0\nu\beta\beta$ decay via light Majorana neutrino exchange was $T^{0\nu}_{1/2}>5\times10^{21}$~y at $90$~\% C.L.~\cite{Arnold:1996wp}.
In recent years the study of the $\beta\beta$ decay of $^{116}$Cd continued using CdWO$_{4}$ scintillator crystals.
The Solotvina experiment~\cite{PhysRevC.68.035501} employed $330$~g of crystals for a total exposure of $0.4$~kg$\times$y and reported a measurement of the $2\nu\beta\beta$ half-life corresponding to $T^{2\nu}_{1/2} = [2.9 \pm 0.06 (\mbox{stat.}) ^{+0.4} _{-0.3} (\mbox{syst.})]\times 10^{19}$~y. The lower limit on the $0\nu\beta\beta$ half-life corresponds to $T^{0\nu}_{1/2} > 1.7 \times 10^{23}$~y at $90$~\% C.L..
The Aurora experiment~\cite{Danevich:2016eot} employs 1.16~kg of crystals enriched to 82~\% in $^{116}$Cd. With an exposure of 0.4~kg$\times$y the measured half-life of the $2\nu\beta\beta$ decay is $T^{2\nu}_{1/2} = [2.62 \pm 0.02 (\mbox{stat.}) \pm 0.14 (\mbox{syst.})]\times 10^{19}$~y. The limit on the $0\nu\beta\beta$ is $T^{0\nu}_{1/2} > 1.9 \times 10^{23}$~y at 90~\%~C.L. and has been obtained with an exposure of 0.3~kg$\times$y.
This paper describe the analysis of the $^{116}$Cd sample installed in the NEMO-3 detector. The NEMO-3 detector is introduced in Section \ref{sec:nemo}. The results of the measurement of the different background components are presented in Section \ref{sec:bkg}. The measurement of the $2\nu\beta\beta$ half-life is summarised in Section \ref{sec:2nu}. The results of the search for the $0\nu\beta\beta$ decay are summarised in Section \ref{sec:0nu}.
%
\section{The NEMO-3 detector \label{sec:nemo}}
%
The NEMO-3 detector performed precise measurements of two neutrino double-beta decay and searched for neutrinoless double-beta decay in seven isotopes, among which were $^{100}$Mo ($\sim7$~kg)~\cite{Arnold:2015wpy} and $^{82}$Se ($\sim1$~kg)~\cite{PhysRevLett.95.182302}.
Smaller amounts of other isotopes such as $^{130}$Te, $^{116}$Cd, $^{150}$Nd, $^{96}$Zr and $^{48}$Ca were also investigated~\cite{PhysRevC.80.032501,PhysRevLett.107.062504, Argyriades2010168, PhysRevD.93.112008, ::2016dpe}.
The detector~\cite{Arnold:2004xq} had a cylindrical geometry and was divided into 20 sectors. 
Each sector consisted of source planes containing the $\beta\beta$ emitting isotopes surrounded by a tracker and a calorimeter. 
The detector was installed in the \textit{Laboratoire Souterrain de Modane} (LSM), under a rock overburden of 4800 m.w.e. to shield against cosmic rays.
The $\beta\beta$ events were emitted from thin central source foils (40-60 mg/cm$^{2}$) suspended between two concentric cylindrical tracking volumes. 
The tracker was composed of 6180 Geiger cells operating in a gas mixture comprising helium with $4$~\% ethanol quencher, $1$~\% argon and $0.15$~\% water vapour.
The tracker allowed the reconstruction of the decay vertices on the foil with an average resolution of $0.5$~mm on the transversal plane ($xy$) and $8.0$~mm on the longitudinal axis ($z$) for $1$~MeV electrons.
The tracking volume was surrounded by a segmented calorimeter made of 1940 large blocks of plastic scintillator coupled to low radioactivity 5'' and 3'' PMTs. 
The calorimeter provided a timing resolution of $\sigma = 250$~ps while the energy resolution was $\sigma_{E}/E = 5.8\%/\sqrt{E(\mbox{MeV})}$ for the scintillator equipped with 5'' PMTs, and $\sigma_{E}/E = 7.2\%/\sqrt{E(\mbox{MeV})}$ for the scintillator equipped with 3'' PMTs.
The detector was immersed in a 25~G magnetic field to enhance charged particle identification and was shielded from external gamma rays by 19~cm of low activity iron and from neutrons by 30~cm of borated water.
After one year of data taking, an air-tight tent surrounding the detector and a radon-free air flushing facility were installed to reduce the radon contamination in the tracking chamber by a factor of six.
The detector energy calibration has been performed by simultaneously introducing 60 radioactive sources in the detector through 20 calibration tubes located near the source foils.
The absolute energy scale calibration was performed every three weeks with $^{207}$Bi sources which provide internal conversion electrons of $482$~keV and $976$~keV. 
%
%
The rare $1682$~keV internal conversion electron peak of $^{207}$Bi was used to determine the systematic uncertainty on the energy scale which was found to be within $0.2$~\% for $99$~\% of the PMTs. 
The energy scale was also tested with the end-point of the $\beta$ spectrum of $^{90}$Y sources ($Q_{\beta}=2.28$~MeV) and of $^{214}$Bi ($Q_{\beta}=3.27$~MeV) from $^{222}$Rn in the tracker volume. 
The linearity of the PMTs was verified regularly with light injection tests. Deviation from the linear response was found to be within $1$~\% for energies $<4$~MeV. 
The same light injection system is also used to survey the relative gain and time variation of individual PMTs twice a day.
PMTs that show a gain variation above $5$~\% compared to a linear interpolation between two successive absolute calibrations with $^{207}$Bi are excluded from the analysis.
Monte Carlo (MC) simulations are performed using the DECAY0 event generator \cite{Ponkratenko:2000um} with a GEANT3-based \cite{brun1987geant3} detector simulation.
The main feature of NEMO-3 was its unique capability to fully reconstruct the kinematics of the events. 
This allowed for topological selection of the events to be performed among the final states of interest. This helps to reduce backgrounds and to discriminate between different mechanisms beyond the neutrinoless double-beta decay~\cite{Arnold:2010tu}.
A total of $440$~g of metallic cadmium was placed in one sector of the NEMO-3 detector.
The cadmium sample was enriched in $^{116}$Cd by the centrifugation separation method.
An average enrichment of $(93.2\pm0.2)$~\% was achieved yielding a total amount of $(410 \pm 1)$~g of $^{116}$Cd.
%
%
Part of the sample (152~g) was previously measured with the NEMO-2 prototype~\cite{Arnold:1995jz,Arnold:1996wp}.
%
%
The Cd was distributed in five strips $\sim242.3$~cm long and $\sim6.5$~cm wide, and two additional strips $\sim6.3$~cm wide placed on the edges of the sector.
Each strip was composed of one piece or smaller pieces that were glued together. 
The full strip was then glued between two $12$~$\mu$m thick Mylar sheets to provide mechanical strength in the vertical position.
The NEMO-3 detector took data from February 2003 until January 2011.
Only runs with stable detector conditions and good energy calibrations are considered.
The data taking is divided into two phases which correspond to the run period before (Phase 1) and after (Phase 2) the installation of the anti-radon facility.
The total live time from both phases is 5.26~y which corresponds to a total exposure of 2.16~kg$\times$y with $^{116}$Cd.
%
\section{Backgrounds \label{sec:bkg}}
\begin{figure}
\includegraphics[scale=.4]{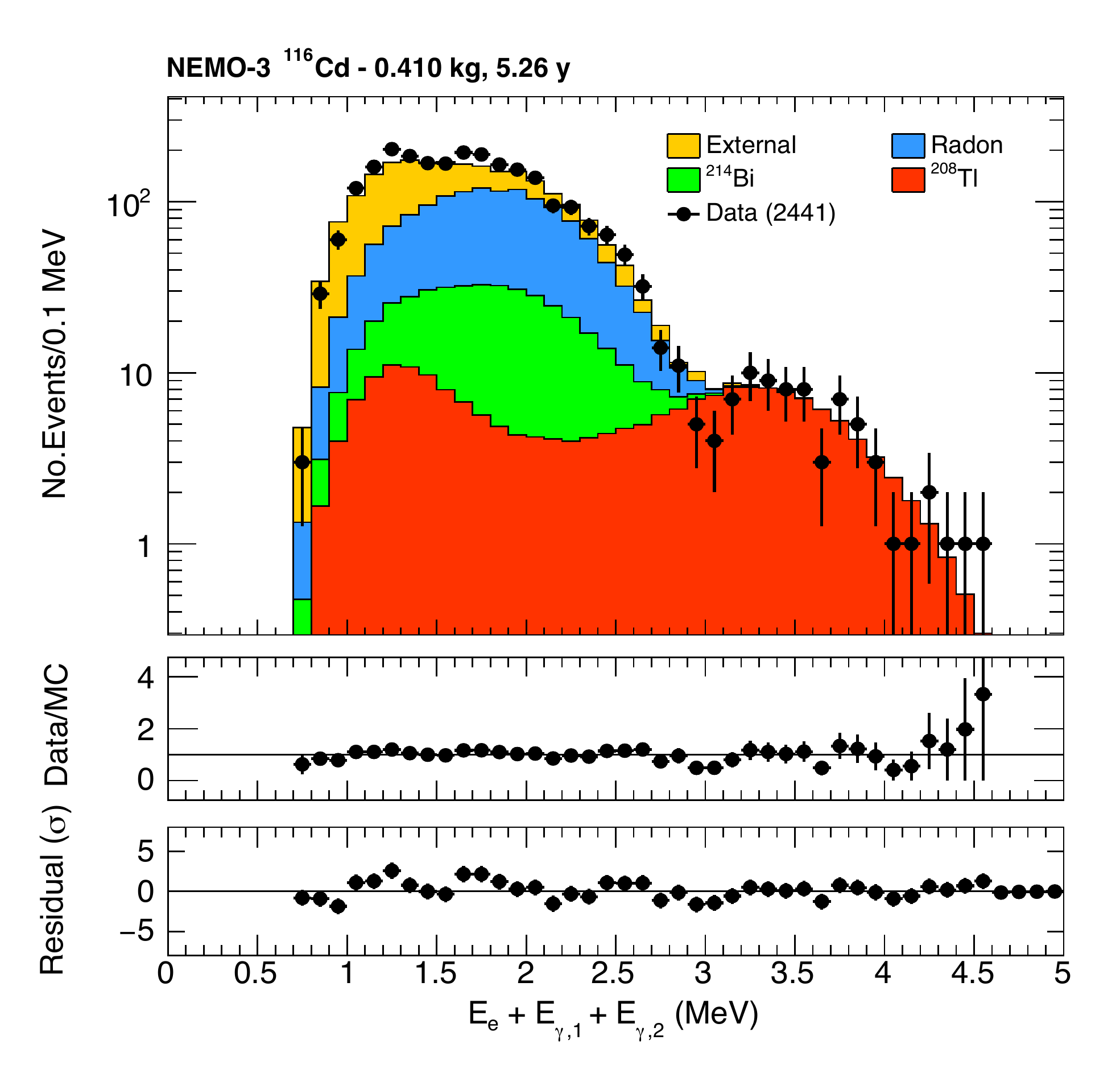}
\caption{Distribution of the total energy in the $1e2\gamma$ channel which provides the most sensitive measure of $^{208}$Tl. The contribution from the external background, radon progeny in the tracking chamber and from $^{214}$Bi are also considered. The ratio of data events to the total Monte Carlo model and the residuals are shown in the bottom panels.
\label{fig:bkg_tl_208}}
\end{figure}

\begin{figure}
\includegraphics[scale=.4]{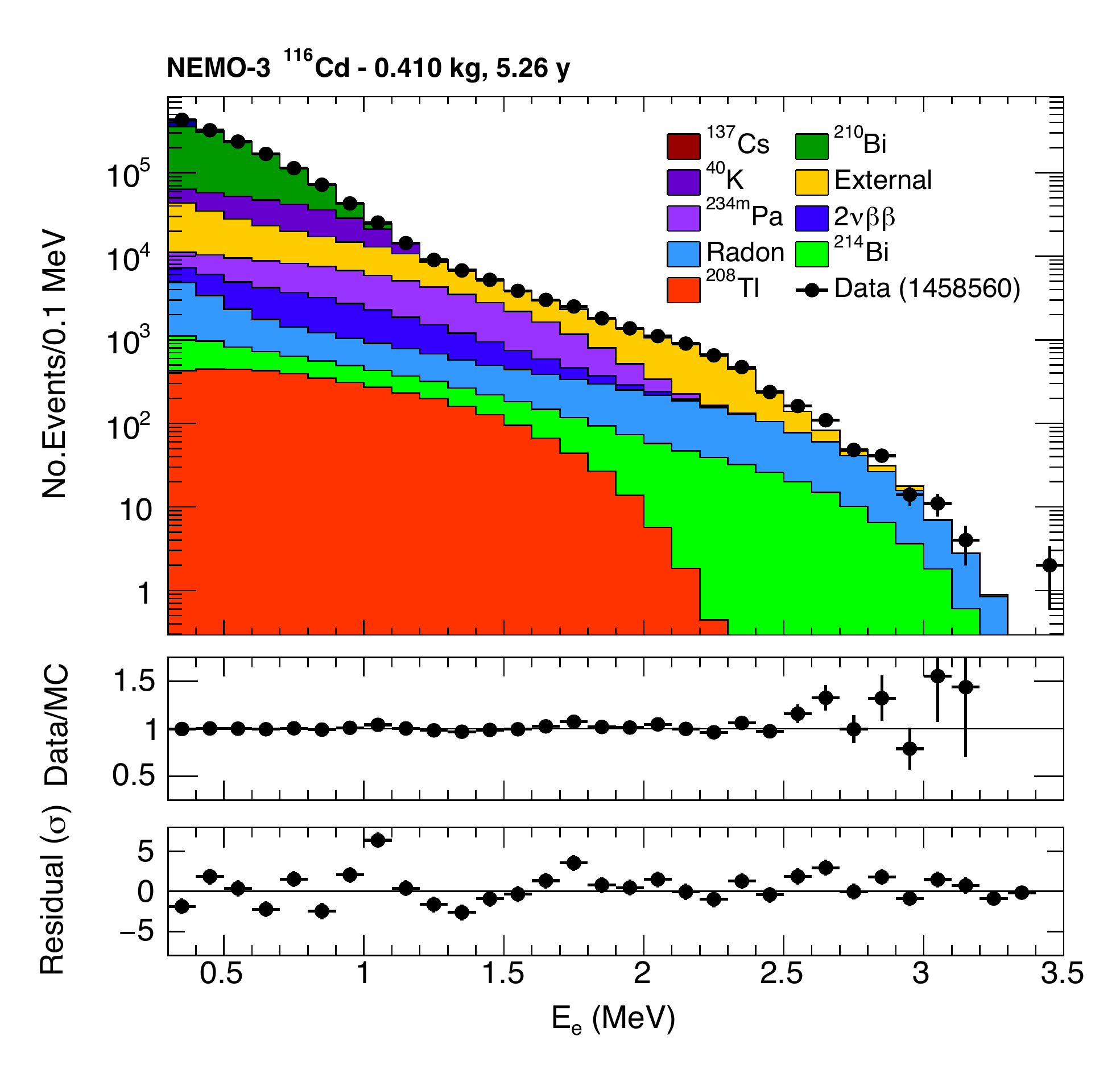}
\caption{Distribution of the electron energy from the $1e$ channel. The dominant contribution at low energies is due to single-$\beta$ emitting isotopes: $^{40}$K, $^{234m}$Pa and $^{210}$Bi. At higher energies the radon progeny in the tracking chamber and the internal $^{214}$Bi become significant. There are sub-leading contributions from other internal and external backgrounds as well as from $2\nu\beta\beta$. The ratio of data events to the total Monte Carlo model and the residuals are shown in the bottom panels.\label{fig:bkg_int}}
\end{figure}

%
The most important backgrounds come from the natural radioactivity of the detector materials due to the presence of long half-life radionuclides, mainly $^{238}$U, $^{232}$Th and their high $Q$-value decay products such as $^{214}$Bi ($Q_{\beta}=3.27$~MeV) and $^{208}$Tl ($Q_{\beta}=4.99$~MeV). 
The presence of $^{222}$Rn emanated from the cavern walls is also particularly troublesome. 
With its relatively long half-life (3.824~days), $^{222}$Rn can diffuse within the detector and deposit $^{214}$Bi in the tracking chamber. 
The internal background originates in the source foil from contamination introduced during isotope enrichment, residual contamination after the $^{116}$Cd purification or during the foil production.
These include $^{238}$U ($^{234m}$Pa) and $^{232}$Th ($^{228}$Ac) decay chains, $^{137}$Cs and $^{40}$K. In particular, $^{228}$Ac gives rise to $^{220}$Rn with subsequent generation of $^{208}$Tl.
The single-$\beta$ emitters or external $\gamma$-rays can mimic two-electron events by double Compton scattering or combined Compton-M{\o}ller scattering in the foil containing the $\beta\beta$ emitter isotope.
Since the NEMO-3 detector was capable of identifying different types of particles and event topologies, it has been possible to study the backgrounds by combining tracking, calorimetric and timing information  in different channels. 
Activities of different background components are then obtained by adjusting Monte Carlo distributions to the data in different channels via a binned log-likelihood fit.
The NEMO-3 backgrounds are evaluated detector-wise in~\cite{Argyriades:2009vq}.
In order to provide a more accurate modelling, a dedicated analysis of the backgrounds in the sector containing the $^{116}$Cd foils is performed.
%
%
%
\begin{figure}
\includegraphics[scale=.4]{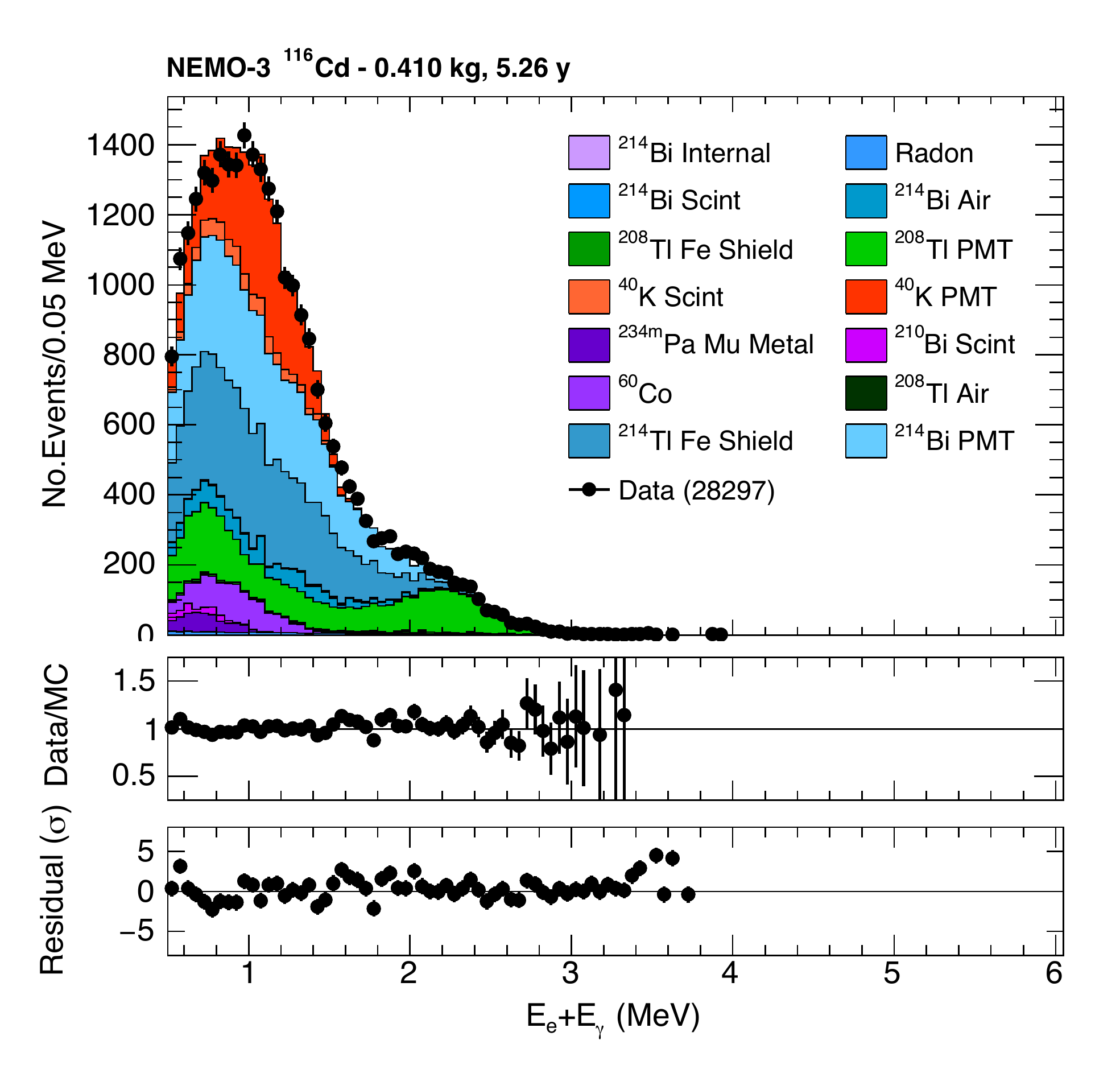}
\caption{Distribution of the total energy in the external $\gamma$-$e$ channel. The external background is modelled with $^{40}$K, $^{214}$Bi, $^{228}$Ac/$^{212}$Bi/$^{208}$Tl, $^{210}$Bi, $^{60}$Co and $^{234m}$Pa in the PMT, scintillator surface, iron structure, iron shield and the air surrounding the detector. The ratio of data events to the total Monte Carlo model and the residuals are shown in the bottom panels. \label{fig:bkg_externals}}
\end{figure}
\\ The backgrounds coming from the outer part of the detector and from the external $\gamma$-ray flux are determined using both external $\gamma$-$e$ and crossing-electron events. 
The external $\gamma$-$e$ events are generated by an isolated calorimeter hit consistent with a $\gamma$-ray and an electron produced by Compton scattering in the source foil whose track is associated with a different calorimeter hit. 
The timing of the calorimeter hits must be consistent with an external $\gamma$-ray hitting the first calorimeter before producing the electron in the foil.
The energy distribution of the $\gamma$-$e$ events is shown in Fig.~\ref{fig:bkg_externals}. 
The crossing-electron events are generated by an electron created by Compton scattering in the calorimeter whose track crosses the entire tracking chamber before hitting a second calorimeter. 
The time-of-flight and the curvature of the track must be consistent with a crossing electron.
Both the $\gamma$-$e$ and crossing-electron events are required to have a vertex on the $^{116}$Cd foil.
The external background is modelled with $^{40}$K, $^{214}$Bi, $^{228}$Ac/$^{212}$Bi/$^{208}$Tl, $^{210}$Bi, $^{60}$Co and $^{234m}$Pa in the PMT, scintillator surface, iron structure, iron shield and the air surrounding the detector. 
%
%
The deviation between the measurement of the external background in the $^{116}$Cd sector and the results from~\cite{Argyriades:2009vq}, which correspond to the average external background measured across the entire detector is $\pm15$~\%. The value is obtained by summing in quadrature the deviation observed for each component of the background model.
%
%
%
\begin{figure}
\includegraphics[scale=.4]{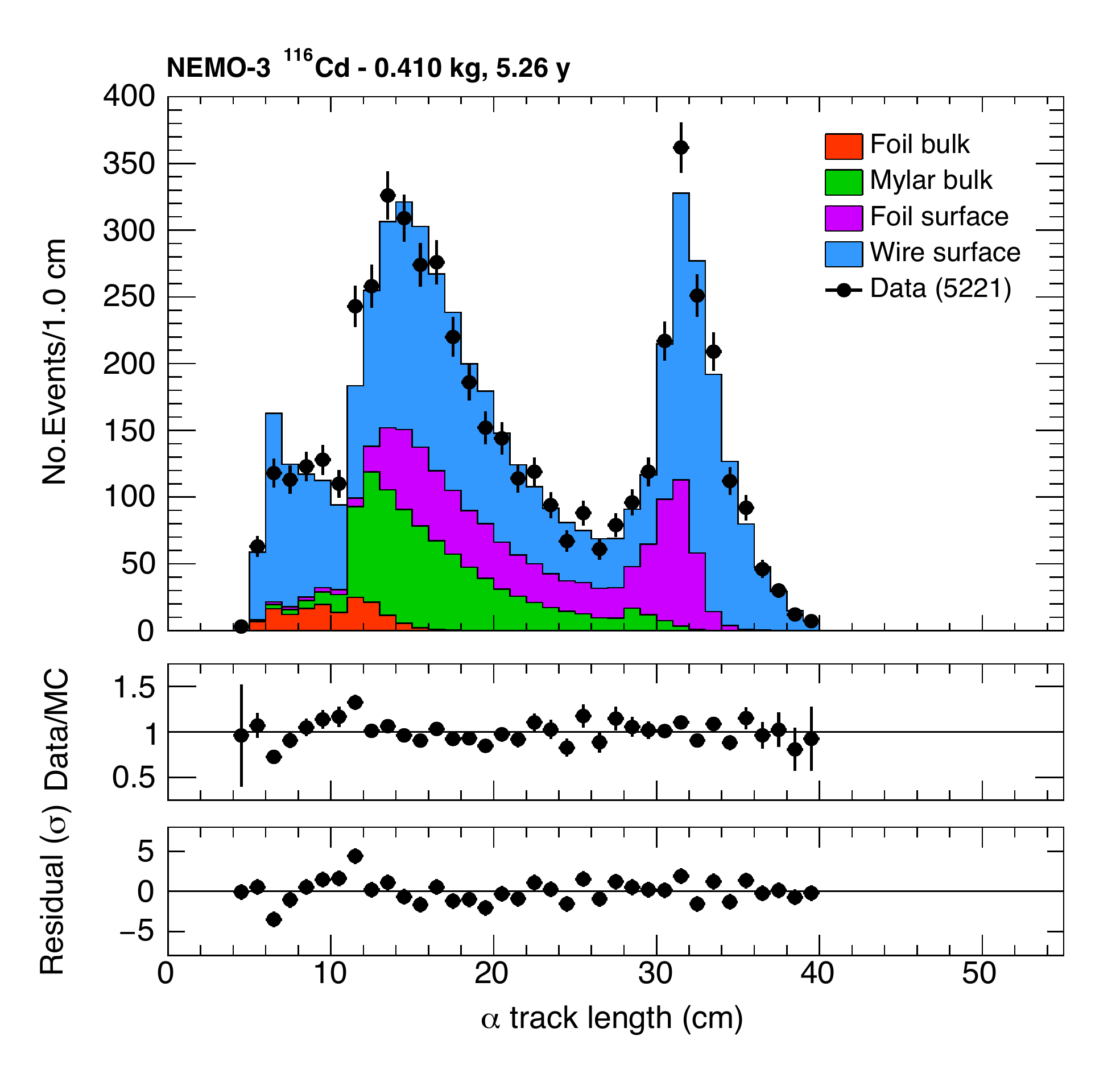}
\caption{Distribution of the delayed $\alpha$ track length in the $1e1\alpha$ channel. The contribution of the radon progeny from the wires of the tracking chamber, the Mylar films which envelop the foils, and the foil surface are shown. The small contribution at small track lengths is consistent with $^{214}$Bi contained in the source foil bulk. The ratio of data events to the total Monte Carlo model and the residuals are shown in the bottom panels. \label{fig:bkg_bi_214}}
\end{figure}
\\ The background coming from internal $^{214}$Bi contamination and from $^{222}$Rn gas in the tracking chamber is determined studying $1e1\alpha$ events produced by the $^{214}$Bi$\rightarrow$$^{214}$Po cascade of the $^{222}$Rn chain.
An $\alpha$ is defined by a short track with one or more delayed Geiger hits in the proximity of the electron vertex. 
The coincidence between the electron and the $\alpha$ event is delayed by the $^{214}$Po half-life of 163.6~$\mu$s~\cite{ShamsuzzohaBasunia2014561} which allows a clean sample of $^{214}$Bi to be obtained. 
Given the high ionisation of $\alpha$ particles, their track length provides the best discrimination of $^{214}$Bi contamination of different origins. The distribution of the delayed $\alpha$ track length of Fig.~\ref{fig:bkg_bi_214} shows evidence of radon contamination on the tracker wires, on the surface of the foil and from the foil itself. 
The $1e1\gamma$ channel is also sensitive to $^{214}$Bi from internal contamination and to $^{222}$Rn in the tracker since $^{214}$Bi usually decays to an excited state of $^{214}$Po with subsequent $\gamma$ emission.
The contamination of $^{214}$Bi and $^{222}$Rn measured in the $1e1\gamma$ and $1e1\alpha$ channels are compatible within $2.0$~\% and $2.6$~\% respectively.
%
%
The $^{208}$Tl contamination in the source foil is determined through its decay chain which consists of a $\beta$ decay accompanied by one or more $\gamma$s from the excited state of $^{208}$Pb~\cite{Martin20071583}. 
Requiring an electron in coincidence with two $\gamma$s provides the most sensitive channel to measure the $^{208}$Tl contamination.
The distribution of the summed electron and $\gamma$ energies in the $1e2\gamma$ channel is shown in Fig.~\ref{fig:bkg_tl_208}.
The distribution is dominated by $^{208}$Tl for summed energies above 3~MeV.
A compatible contamination is observed requiring one electron in coincidence with just one $\gamma$.
%
%
\\ Backgrounds produced by single-$\beta$ emitters are determined with single electrons originating from the $^{116}$Cd foils.
%
%
%
\begin{figure}
\includegraphics[scale=.4]{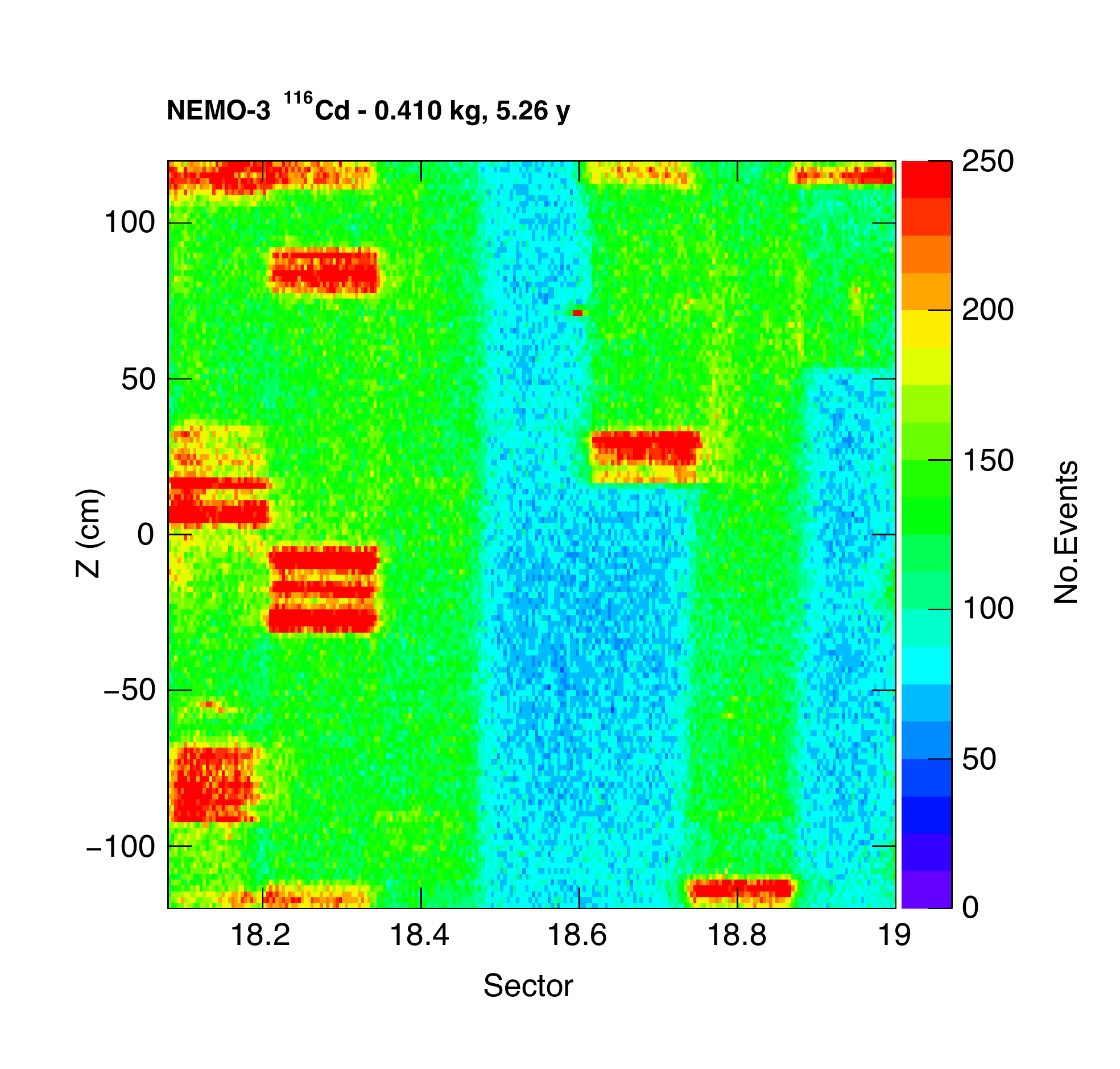}
\caption{Distribution of the track vertices on the source foil in the $1e$ channel. Three regions are defined based on the number of recorded events. The high activity regions, consistent with an excess of $^{234m}$Pa on the surface of the foil, are rejected from the analysis. \label{fig:bkg_foil_region}}
\end{figure}
A non-uniform distribution of the vertex location of the electron tracks on the foil surface is observed as shown in Fig.~\ref{fig:bkg_foil_region} with an arbitrary bin width.
Three foil regions with different levels of activity appear evident: a high activity region with $\gtrsim200$~events/bin, a medium activity region with $[100 - 200]$~events/bin and a low activity region with $\lesssim100$ events/bin.
No evidence of high activity spots is observed on the vertex map of events selected in the $1e1\gamma$ channel, suggesting their origin is due to a contamination from single  $\beta$-emitters.
Since these regions present a well defined shape in the source foil plane, they are defined through a set of rectangular cuts based on the number of events per bin.
The medium and low activity region represent $89$~\% of the foil surface. The background model in these regions is consistent with contamination from  $^{40}$K, $^{234m}$Pa, $^{210}$Bi and $^{137}$Cs as shown in Fig.~\ref{fig:bkg_int}. 
The magnitude of the background activity in these regions differs by about a factor of two.
The high activity regions represent $11$~\% of the foil surface. The background model is consistent with an excess of $^{234m}$Pa, most likely due to the excess of glue used to connect small pieces of the $^{116}$Cd foil to form the full strip.
The excess of $^{234m}$Pa is of a factor of about 20 and 40 with respect to the medium and low activity regions respectively. 
The high activity regions are rejected from the analysis.
\\ The activity of the different background contributions and their impact to the measurement of the $2\nu\beta\beta$ half-life and the search for $0\nu\beta\beta$ decay is summarised in Tab.~\ref{tab:bkg}.
\begin{table}
\caption{Summary of measured background activities (A), expected number of background events in the two-electrons channel and number of $2\nu\beta\beta$ events (N). The values for the internal background are given for the low and medium activity regions. 
Secular equilibrium is assumed between $^{214}$Bi and $^{214}$Pb. The same is done between $^{208}$Tl, $^{228}$Ac, and $^{212}$Bi, where the branching ratio of $35.94$~\% is taken into account. The quoted uncertainties are statistical only. \label{tab:bkg}}
 \begin{ruledtabular}
 \begin{tabular}{llcc}
Contributions                    & Activity region   & A (mBq/kg)    & N            \\ \hline
\multicolumn{2}{l}{$^{208}$Tl}              & 0.13$\pm$0.03 & 19$\pm$2     \\
\multicolumn{2}{l}{$^{214}$Bi}              & 0.4$\pm$0.1   & 30$\pm$5     \\
\multirow{2}{*}{$^{210}$Bi}       & Low	    & 140$\pm$1     & 10$\pm$4     \\ 
                                  & Medium  & 337$\pm$1     & 23$\pm$10    \\
\multirow{2}{*}{$^{40}$K}         & Low	    & 12.9$\pm$0.5  & 9.0$\pm$0.5  \\                                                                                      
                                  & Medium  & 23.7$\pm$0.5  & 26$\pm$1     \\
\multirow{2}{*}{$^{234m}$Pa}      & Low	    & 2.7$\pm$0.5   & 28$\pm$5     \\                                                                                      
                                  & Medium  & 5.1$\pm$0.5   & 73$\pm$7     \\
\multicolumn{2}{l}{Externals}               & $-$           & 136$\pm$14   \\
\multicolumn{2}{l}{Radon}                   & $-$           & 49$\pm$2     \\
\multicolumn{2}{l}{Total background}        & $-$           & 402$\pm$19   \\
\multicolumn{2}{l}{$2\nu\beta\beta$}        & $-$           & 4968$\pm$74 \\ \hline
\multicolumn{2}{l}{Data}                    & $-$           & 5368
 \end{tabular}
 \end{ruledtabular}
 \end{table}
%
\section{Two neutrino double-beta decay \label{sec:2nu}}
\begin{figure*}
\includegraphics[scale=.4]{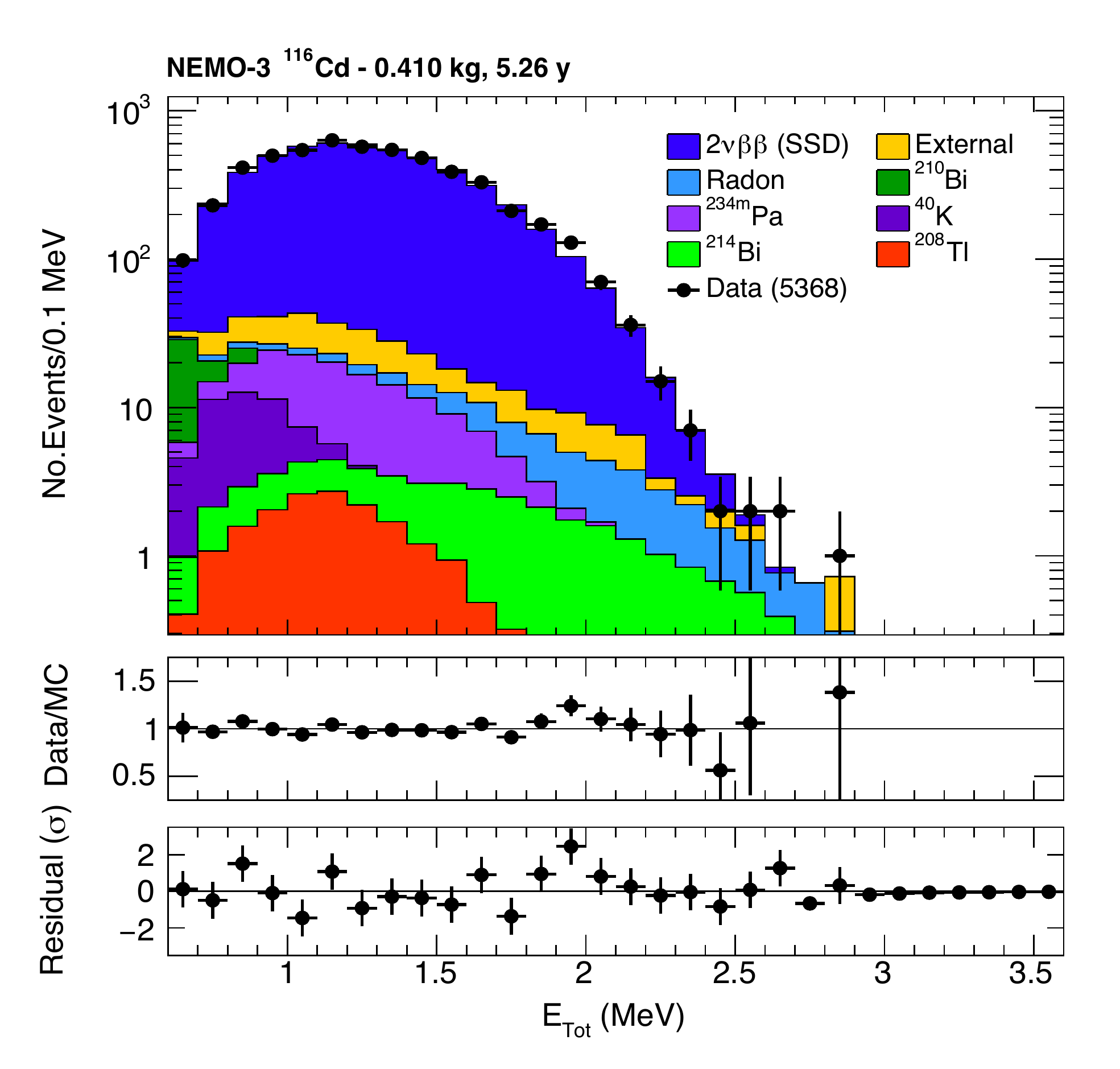}
\includegraphics[scale=.4]{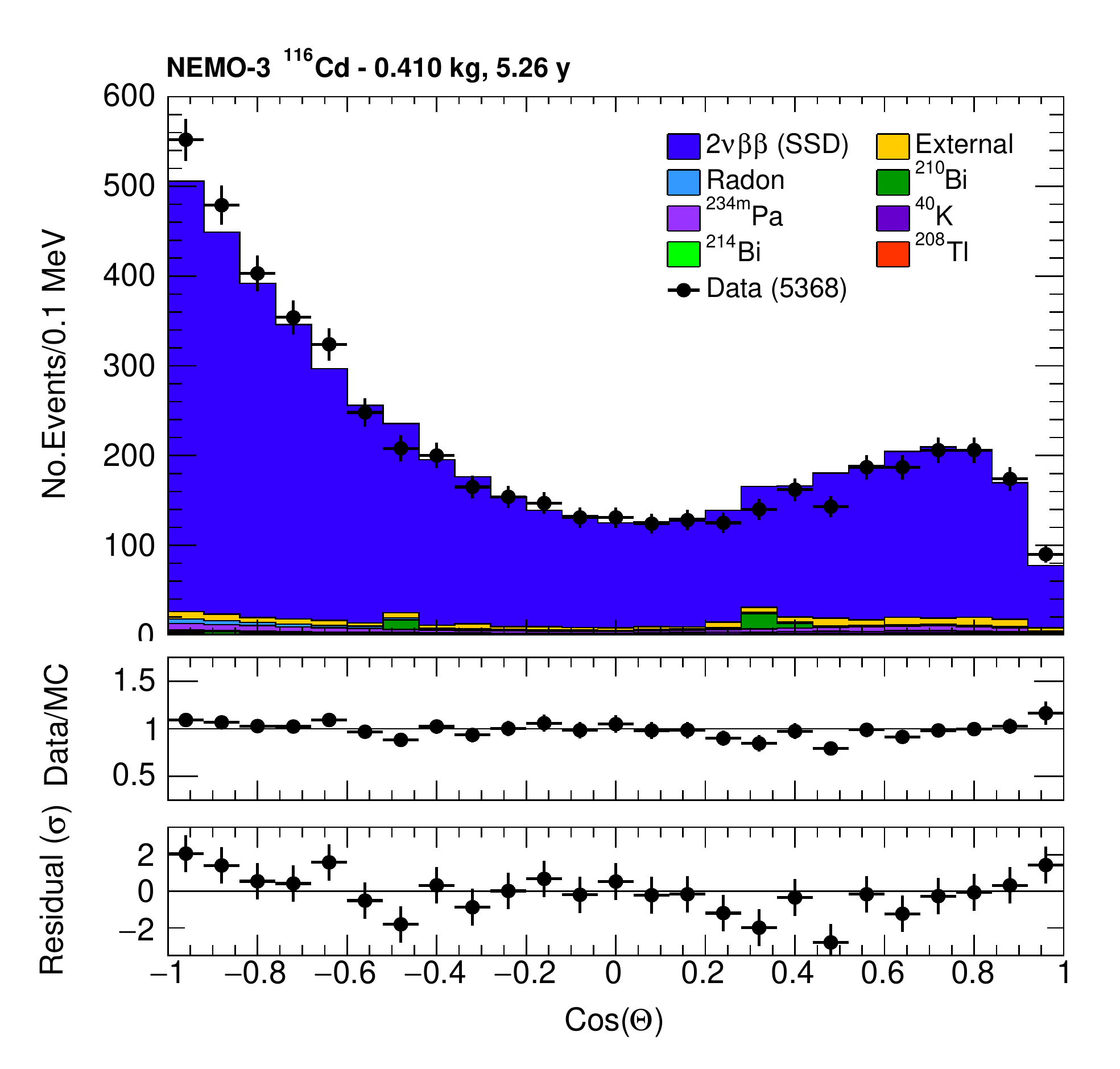}
\caption{Distribution of the total electron energy (left) and the angle between the two electron tracks (right) in the two-electron channel. Background contributions are due to external events, radon progeny in the tracker chamber, internal $^{208}$Tl, $^{214}$Bi and single-$\beta$ emitters. The ratio of data events to the total Monte Carlo model and the residuals are shown in the bottom panels.\label{fig:2nubb}}
\end{figure*}

\begin{figure*}
\includegraphics[scale=.4]{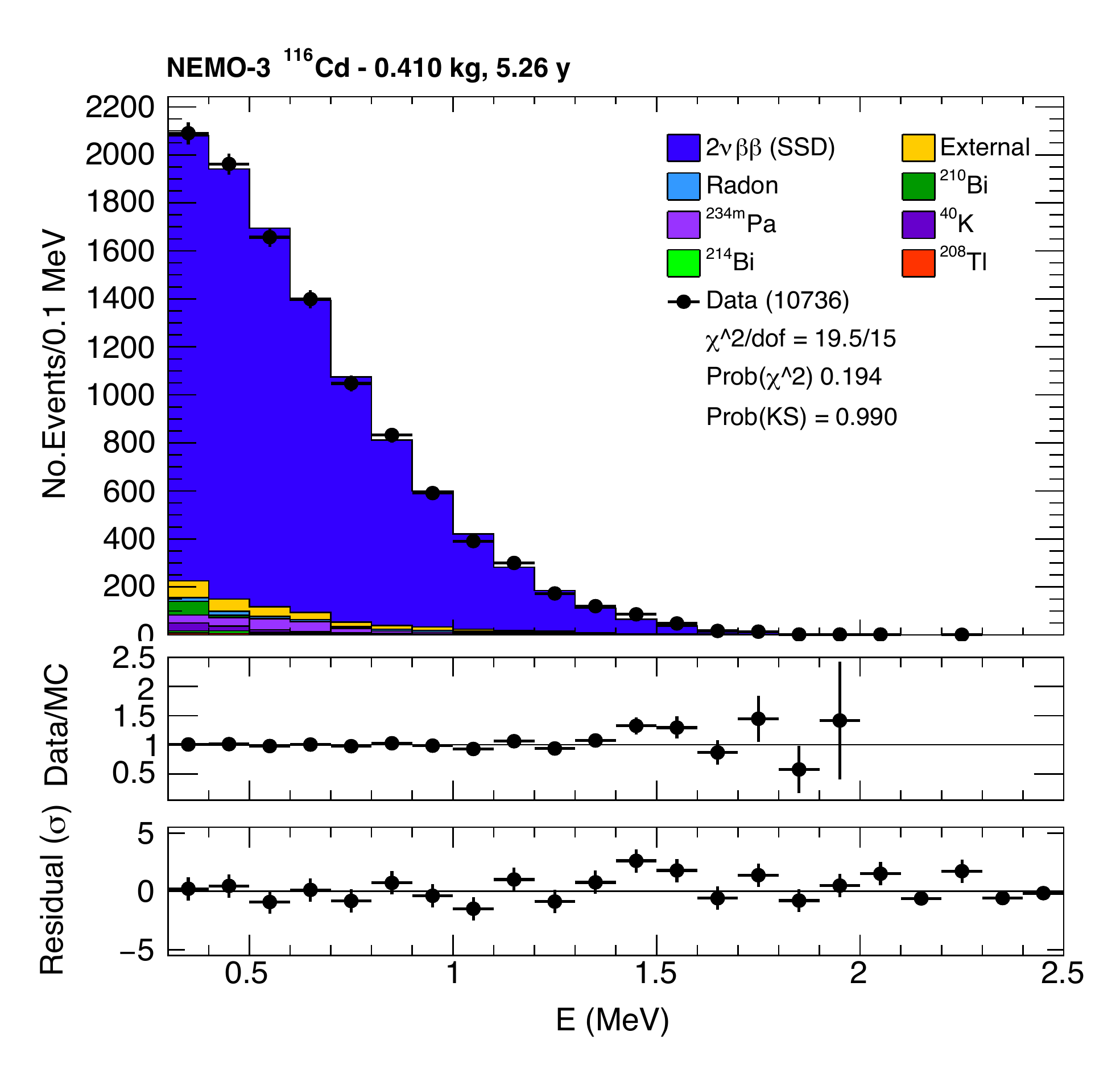}
\includegraphics[scale=.4]{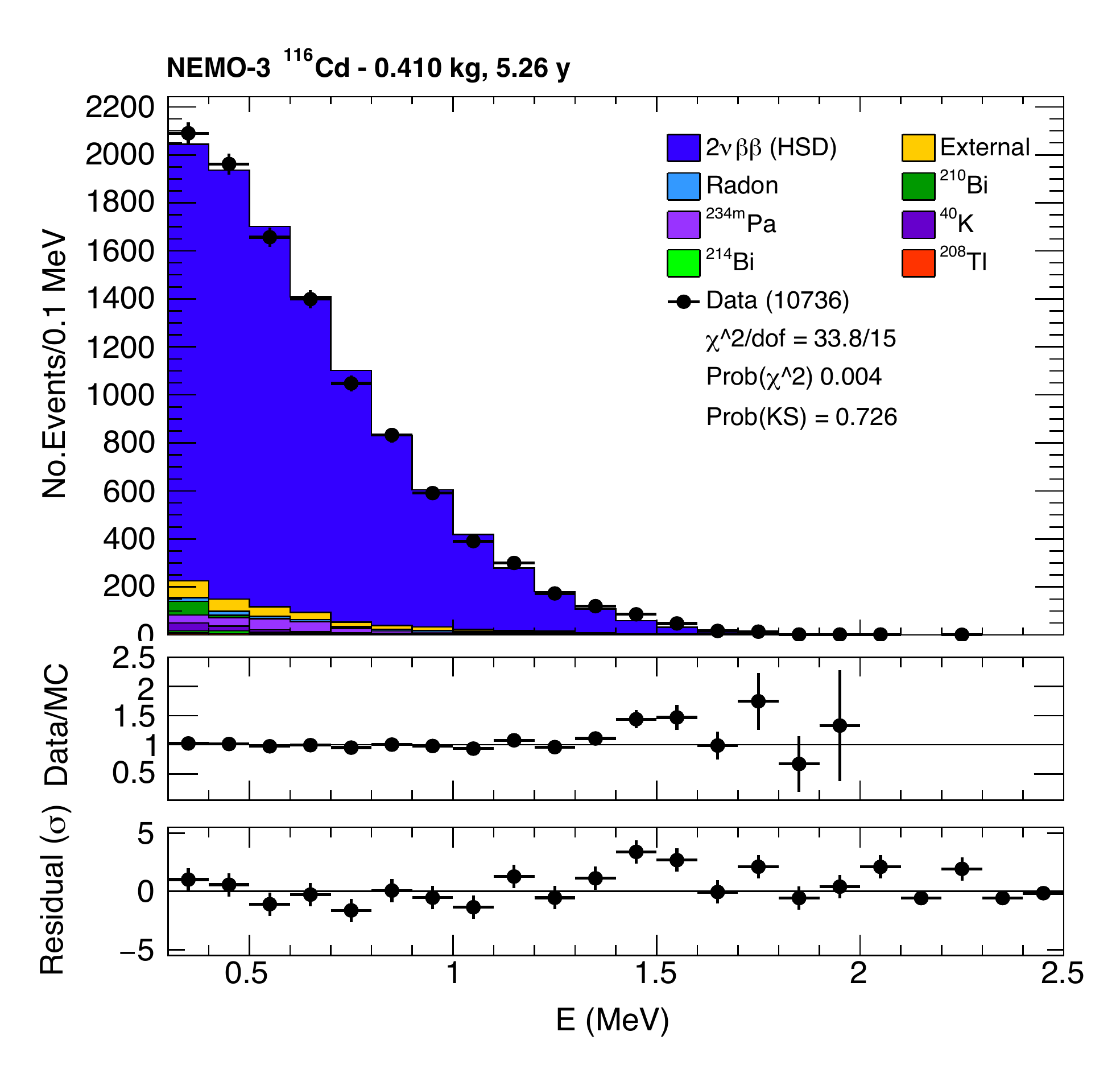}
\caption{Distribution of the single electron energy in the two electron channel under the SSD (left) and the HSD (right) hypothesis. Background contributions are due to external events, radon progeny in the tracker chamber, internal $^{208}$Tl, $^{214}$Bi and single-$\beta$ emitters. The ratio of data events to the total Monte Carlo model and the residuals are shown in the bottom panels. In the SSD (HSD) hypothesis the p-value of the Kolmogorov-Smirnov test is 0.990 (0.726) while the p-value of the $\chi^{2}$ test is 0.193 (0.004). The results of the tests are in favour of the SSD hypothesis but there is not enough sensitivity to exclude the HSD hypothesis. \label{fig:SSD_HSD}}
\end{figure*}

\begin{figure}
\includegraphics[scale=.4]{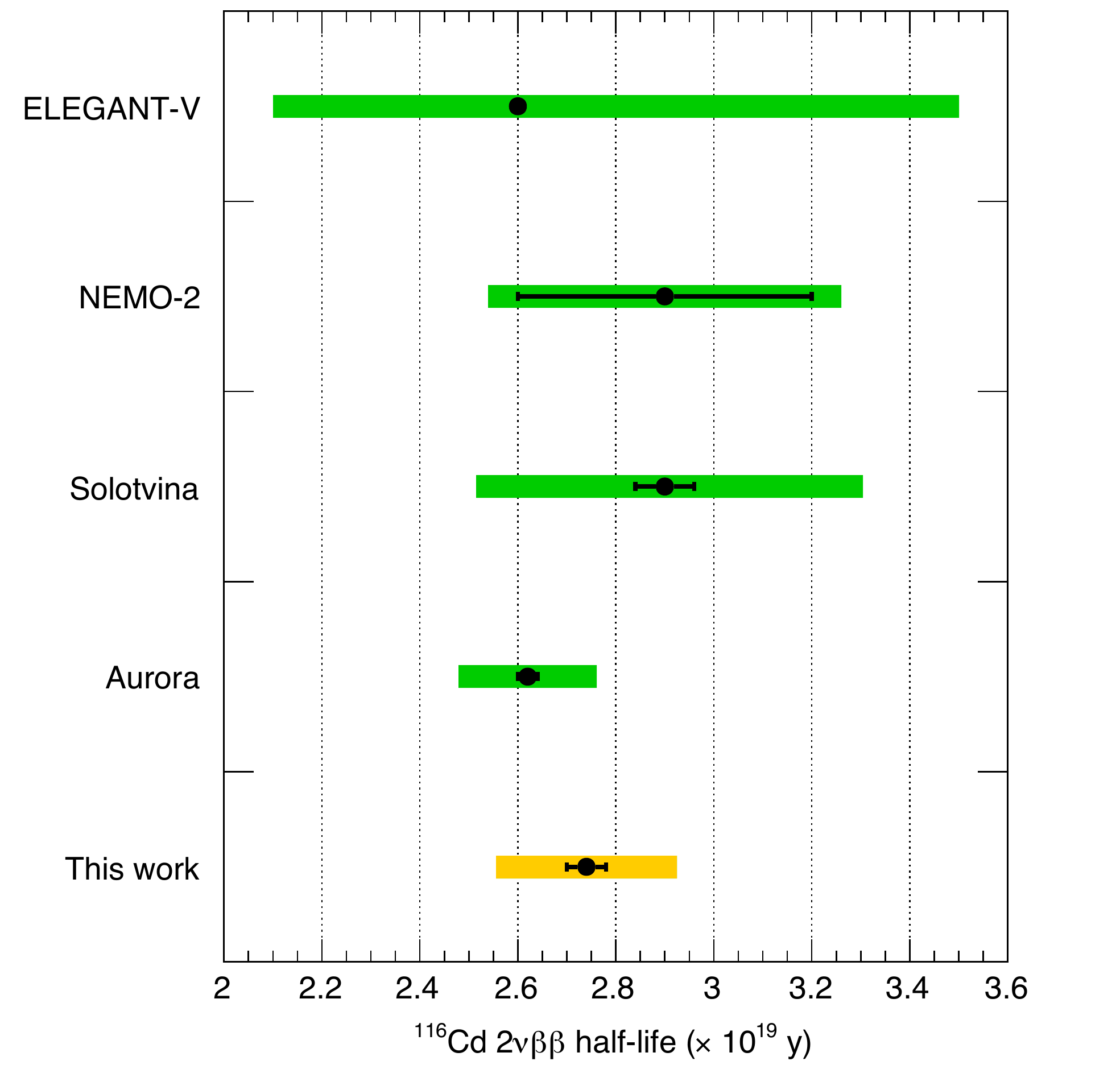}
\caption{
Comparison of the $2\nu\beta\beta$ half-life of $^{116}$Cd measured in this work with respect to previous measurement from Aurora~\cite{Danevich:2016eot}, Solotvina~\cite{PhysRevC.68.035501}, NEMO-2~\cite{PhysRevC.81.035501} and ELEGANT-V~\cite{Kume}. The black bars show the statistical uncertainty while the coloured bars shows the total statistical and systematical error combined in quadrature.
\label{fig:comp}}
\end{figure}

%
The best sensitivity to measure the $2\nu\beta\beta$ decay rate is obtained in the $2e^{-}$ channel.
Candidate events must have exactly two electrons with reconstructed vertices inside the $^{116}$Cd foils. 
The separation between each individually reconstructed vertex must be less than $4$~cm radially and $8$~cm longitudinally to ensure that the electrons originate from a common vertex. 
Electron tracks with vertices lying on the high activity regions of the foil which contain an excess of $^{234m}$Pa are rejected in order to improve the signal to background ratio. 
Each of the electron tracks must be at least $50$~cm long, have a hit in the first layer of Geiger cells, and be associated to separate scintillator blocks. 
%
%
The impact of the tracks must be on the front face of the scintillator block to ensure optimal energy reconstruction. 
%
%
The energy of each electron is required to be at least $300$~keV, as the event rate is dominated by background decays below this threshold.
No delayed tracker hits are allowed within $15$~cm of the electron vertices in order to reduce contamination from $^{214}$Bi. 
To further improve rejection of $^{214}$Bi, prompt Geiger hits unassociated to the electron tracks are not allowed on the opposite side of the foil if the electron tracks are on the same side of the foil. 
Additional, isolated calorimeter hits which do not have a track associated to them are allowed if the energies are less than 200~keV. 
This allows for a low level of calorimeter noise in the event, which is not simulated and would therefore introduce an efficiency bias if removed. 
The timing and the path length of the electrons must be consistent with an origin from the common vertex on the $^{116}$Cd source foil.
A total of $5368$ data events pass these selection criteria.
The expected number of events from the $2\nu\beta\beta$ decay of $^{116}$Cd is determined using a log-likelihood fit to the sum of the two electron energy distribution dividing the sample into phase 1 and phase 2, and into the medium and low internal background regions of the foil.
The background activities are constrained through Gaussian parameters to the values and uncertainties measured in Sec.~\ref{sec:bkg} and summarised in Tab.~\ref{tab:bkg}.
%
%
The distribution of the total electron energy and of the angle between the two electron tracks are shown in Fig.~\ref{fig:2nubb}. The MC model at the best fit normalisation is compared to the data with a $\chi^{2}$ test, which provide $\chi^{2}/d.o.f. = 17.8/17$ for the total electron energy distribution and $\chi^{2}/d.o.f. = 29.4/24$ for the angular distribution.
The $2\nu\beta\beta$ decay proceeds through the $J^{\pi}=1^+$ states of the intermediate nucleus via virtual $\beta$ transitions which occur in two steps. 
The first step involves transitions connecting the ground state of the initial nucleus with the $1^+$ states of the intermediate nucleus. 
The second step involves transitions connecting the intermediate $1^+$ states to the ground state of the final nucleus. 
If the two-step process runs exclusively through the first $1^+$ state, the process is said to be single-state dominated (SSD) \cite{Abad84}. 
On the other hand, if the two step process runs through higher excited states, the process is said to be higher-state dominated (HSD).
The subject of nuclear structure interest is whether the $2\nu\beta\beta$-decay nuclear matrix element is dominated by contributions through the lowest state of intermediate nucleus or through higher excited states, in particular through states located in Gamow-Teller resonance region. 
The calculation of the $2\nu\beta\beta$-decay half-lives within the SSD approach indicate the importance of this assumption~\cite{Civ98}.
In~\cite{Sim01,Dom05} it was suggested that this issue can be studied by comparing measured differential characteristics with those calculated within the SSD and the HSD assumptions. 
In particular the single electron energy, an observable available in NEMO-3, is expected to discriminate the two hypotheses.
The $2\nu\beta\beta$ decay of $^{116}$Cd has been found suitable for such an analysis due to a significant dependence of the SSD differential characteristics on the lepton energies in energy denominators~\cite{Dom05}.
Two MC samples are produced, each using one of the above models, and compared to the $2e^{-}$ sample.
Kolmogorov-Smirnov and $\chi^{2}$ tests are performed on the single electron energy distribution shown in Fig.~\ref{fig:SSD_HSD} under both hypotheses.
In the SSD (HSD) hypothesis the p-value of the Kolmogorov-Smirnov test is 0.990 (0.726) while the p-value of the $\chi^{2}$ test is 0.193 (0.004). 
The results of the tests are in favour of the SSD hypothesis but there is not enough sensitivity to exclude the HSD hypothesis.
%
%
The number of $2\nu\beta\beta$ events obtained from the likelihood fit is $N_{2\nu}=(4968\pm74)$ for the SSD hypothesis and $N_{2\nu}=(4966\pm74)$ for the HSD hypothesis, while the total number of background events is $N_{b}=(402\pm19)$.
The number of signal events over the number of total background events is $S/B\simeq12$ with the external background and the $^{234m}$Pa from the source foil being the most dominant contributions. 
The signal selection efficiency estimated from the MC is $\epsilon=(1.8\pm0.1)$~\% for the SSD hypothesis and $\epsilon=(1.9\pm0.1)$~\% for the HSD hypothesis.
The half-life of the $2\nu\beta\beta$ decay is given by:
\begin{equation}
T_{1/2}^{2\nu}= \frac{N_{A} \ln(2)}{W} \cdot \epsilon \cdot \frac{M\times t}{N_{2\nu}}
\end{equation}
where $N_{A}$ is the Avogadro number, $W$ is the atomic weight of $^{116}$Cd, $M$ is the total mass of  $^{116}$Cd in NEMO-3 and $t$ is the total exposure time.
\\ Apart from the statistical uncertainties of the available data sample, the measurement of the $2\nu\beta\beta$ decay half-life is subjected to different sources of systematic uncertainties related to the detector response, the modelling of the $^{116}$Cd source and the measurement of the backgrounds.
%
%
The uncertainty on the $2e^{-}$ reconstruction efficiency is estimated using dedicated runs with two calibrated $^{207}$Bi sources. 
The two conversion electrons emitted by the $^{207}$Bi sources are selected with the same criteria adopted for the $2e^{-}$ channel, except that the common vertex of the two electron tracks must originate from the calibration sources. 
The reconstructed $^{207}$Bi activities agree with the nominal values within 5.5~\% which is assumed as the systematic uncertainty on the detection efficiency $\epsilon$ and directly propagated to the $2\nu\beta\beta$ decay half-life.
%
%
The mass of $^{116}$Cd in NEMO-3 is known within 0.25~\% and it also propagates directly to the $2\nu\beta\beta$ decay half-life.
The systematic uncertainty due to the Monte Carlo modelling of the composition and geometry of the $^{116}$Cd source foils is taken into account measuring the $2\nu\beta\beta$ decay half-life in the medium and low background region of the foil separately.
The uncertainty is quoted from the relative difference, with respect to the half-life obtained in the simultaneous fit, which is $+$2.2~\% for the medium and $-$3.2~\% for the low background region.
%
%
The calorimeter energy scale is known to $\pm$1~\% from periodic surveys of the calorimeter response performed during the life time of the detector with in-situ calibration systems. 
This effect translates into a $\pm$1.2~\% uncertainty on the $2\nu\beta\beta$ decay half-life.
%
%
The difference between the activities measured in the $1e1\alpha$ and $1e1\gamma$ channels are assumed as systematic uncertainty for the $^{214}$Bi and $^{222}$Rn backgrounds. 
The impact on the $2\nu\beta\beta$ decay half-life is of $\pm$0.01~\% and $\pm$0.07~\% for the $^{214}$Bi and $^{222}$Rn respectively.
%
%
The systematic uncertainty on the $^{208}$Tl background is estimated measuring a calibrated $^{232}$U source with events containing one electron and two $\gamma$s.
A discrepancy of 10~\% is observed, which translates to $\pm$0.05~\% on the $2\nu\beta\beta$ decay half-life. 
%
%
The systematic uncertainty on the components of the internal background, $^{40}$K, $^{234m}$Pa and $^{210}$Bi is estimated in the $2e^{-}$ channel, by allowing these contributions to float freely in the likelihood fit. 
The deviation observed on the activities of the internal background is $\pm35$~\%, but the impact on the total number of background events observed remains small.
The effect on the $2\nu\beta\beta$ decay half-life is $\pm1.1$~\%.
%
%
The systematic uncertainty for the external background is $15$~\%, obtained from the difference observed between the measurement of the external background in the $^{116}$Cd sector and the results from~\cite{Argyriades:2009vq}.
The effect on the $2\nu\beta\beta$ decay half-life is $\pm0.45$~\%.
%
%
The different sources of systematic uncertainties discussed in the text are summarised in Tab.~\ref{tab:syst}. 
The total systematic uncertainty on the $2\nu\beta\beta$ decay half-life is obtained by summing in quadrature the different contributions and is found to be $+6.2$~\% and $-6.7$~\%.
\begin{table}
\caption{Summary of systematic uncertainties on the measured half-life for the $2\nu\beta\beta$ decay of $^{116}$Cd.\label{tab:syst}}
 \begin{ruledtabular}
 \begin{tabular}{lr}
Origin                                          & Uncertainty on $T_{1/2}^{2\nu}$ \\ \hline
Electron reconstruction efficiency              & $\pm5.5$~\%       \\
$^{116}$Cd mass                                 & $\pm0.25$~\%      \\
$^{116}$Cd foil modelling                       & $[+2.2,-3.2]$~\% \\
Energy calibration                              & $\pm1.2$~\%       \\
$^{214}$Bi background                           & $\pm0.01$~\%      \\
$^{208}$Tl background                           & $\pm0.05$~\%      \\
Radon background                                & $\pm0.02$~\%      \\
$^{40}$K, $^{234m}$Pa and $^{210}$Bi background & $\pm1.07$~\%      \\
External background                             & $\pm0.45$~\%      \\ \hline
Total                                           & $[+6.2,-6.7]$~\%
\end{tabular}
 \end{ruledtabular}
 \end{table}
%
%
Given the isotope mass and detector exposure of Sec.~\ref{sec:nemo}, the half-life for the $2\nu\beta\beta$ decay of $^{116}$Cd is,
\begin{equation}
 T_{1/2}^{2\nu}=\left[ 2.74\pm0.04\mbox{(stat.)}\pm0.18\mbox{(syst.)} \right] \times10^{19}~\mbox{ y}
\end{equation}
in the SSD hypothesis and,
\begin{equation}
T_{1/2}^{2\nu}=\left[ 2.96\pm0.04\mbox{(stat.)}\pm0.20\mbox{(syst.)} \right] \times10^{19}~\mbox{ y}	
\end{equation}
in the HSD hypothesis.
Fig.~\ref{fig:comp} shows a comparison of the $2\nu\beta\beta$ half-life of $^{116}$Cd measured in this work with respect to previous measurements from Aurora~\cite{Danevich:2016eot}, Solotvina~\cite{PhysRevC.68.035501}, NEMO-2~\cite{PhysRevC.81.035501} and ELEGANT-V~\cite{Kume}.
%
%
%
\section{Neutrinoless double-beta decay \label{sec:0nu}}
\begin{figure}
\includegraphics[scale=.4]{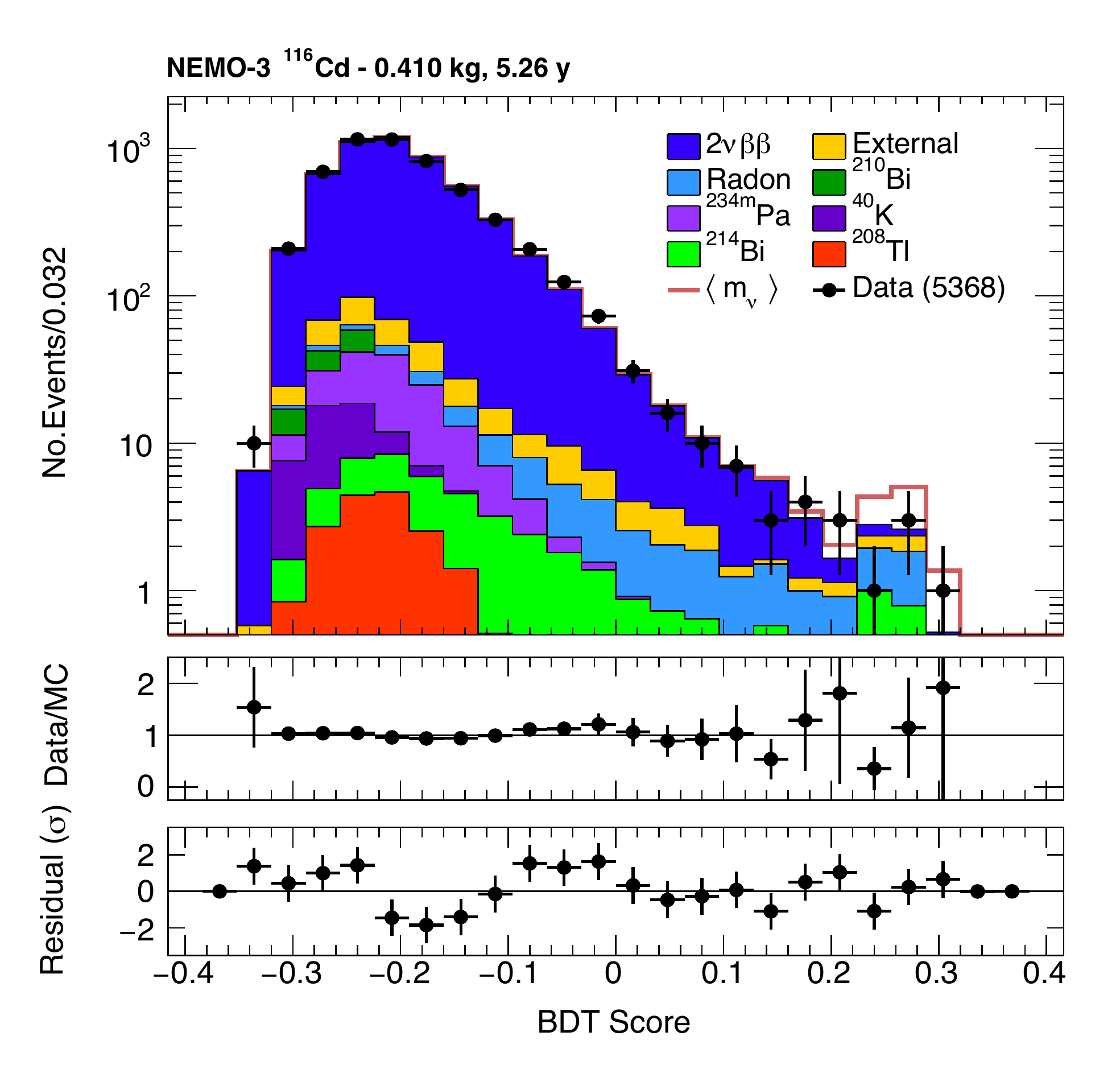}
\caption{Distribution of the output score of the BDT trained for the search of $0\nu\beta\beta$ decays via the exchange of a light Majorana neutrino. The open histogram stacked over the background model shows the 90~\% C.L. limits obtained on the $0\nu\beta\beta$ decay. The ratio of data events to the total Monte Carlo model and the residuals are shown in the bottom panels.\label{fig:BDT}}
\end{figure}

\begin{figure}
\includegraphics[scale=.4]{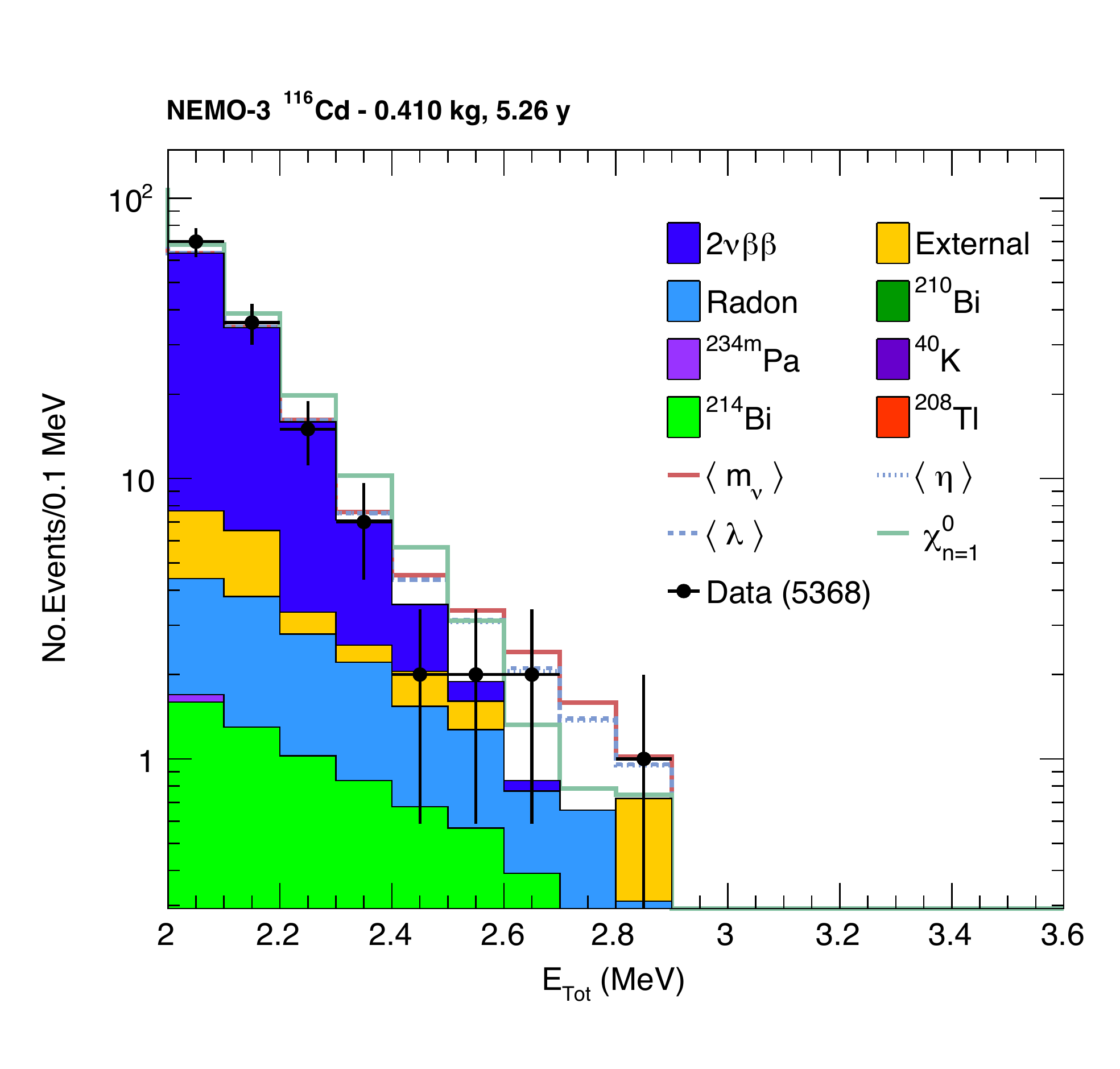}
\caption{Distribution of the total electron energy in two-electron channel above 2~MeV. The four open histograms stacked over the background model show the 90~\% C.L. limits obtained on the $0\nu\beta\beta$ decay for different mechanisms (see text).\label{fig:0nubb}}
\end{figure}

%
The $2e^{-}$ selection described in Sec.~\ref{sec:2nu} is used to search for $0\nu\beta\beta$ decays of $^{116}$Cd to the ground state of $^{116}$Sn.
Different mechanisms expected to produce $0\nu\beta\beta$ decay are investigated.
The $0\nu\beta\beta$ decay through the exchange of a light Majorana neutrino, also referred to as the mass mechanism, is the favoured mechanism. 
%
The half-life of the decay is related to an effective neutrino mass through Eq.~\ref{eq:mm}.
The experimental signature is a peak in the distribution of the total energy of the two electrons near the $Q_{\beta\beta}$ value, which for $^{116}$Cd is ($2813.50\pm0.13$)~keV~\cite{Rahaman2011412}.
The two electrons from the light Majorana neutrino exchange are expected to be selected with an efficiency of ($8.17\pm0.01$)~\%, as estimated from MC simulations.
Within R-parity violating supersymmetric models, the $0\nu\beta\beta$ could proceed through the exchange of a gluino or neutralino~\cite{SUSY}.
To a good approximation the decay kinematics are the same as in the case of the mass mechanism.
In left-right symmetric models, the existence of right-handed weak currents allows the $0\nu\beta\beta$ decay to occur without the helicity flip required by the mass mechanism~\cite{Arnold:2010tu}.
In these models, the amplitude of the $0\nu\beta\beta$ decay does not depend only on the effective neutrino mass but more generally on the coupling between left-handed quarks and right-handed leptons $\langle \eta \rangle$ and on the coupling between right-handed quarks and right-handed leptons $\langle \lambda \rangle$.
Even though these modes provide a total energy distributions peaked at the $Q_{\beta\beta}$, the individual electron kinematics differ from those of the electrons emitted for the mass mechanism.
The different kinematics provide a different signal selection efficiency of ($7.51\pm0.01$)~\% and ($4.16\pm0.01$)~\% for the  $\langle \eta \rangle$ and $\langle \lambda \rangle$ terms respectively, as estimated from the MC simulation.
The lower signal efficiency observed for these models with respect to the mass mechanism is due to smaller opening angle among the two electron tracks. The $\langle \lambda \rangle$ is further suppressed by a significant asymmetry expected between the energies of the electrons.
Finally, $0\nu\beta\beta$ might also proceed via the emission of one or more majorons, weakly interacting bosons present in many GUT theories~\cite{Bamert:1994hb}.
Since the available energy is shared among three or four particles, the signal signature does not correspond to a peak at the $Q_{\beta\beta}$ but rather to a continuous spectrum.
The spectral shape for majoron emission depends on a spectral index $n$.
Decays with larger $n$ have broader summed energy peaked at lower values and are thus more difficult to disentangle from the $2\nu\beta\beta$ decay and other backgrounds.
Given the relatively low statistics and high level of background at low energies, this work considers only the emission of one majoron with spectral index $n=1$ which provides a signal selection efficiency of ($5.55\pm0.01$)~\% as estimated from MC simulation.
While $0\nu\beta\beta$ decay might result from the interference of several mechanisms, in this work they are investigated separately assuming no interference.
\\ The unique feature of NEMO-3 among competing techniques is the capability of reconstructing the full topology of the final state in each event.
This capability might allow one to distinguish between the various underlying $0\nu\beta\beta$ decay mechanisms. 
Furthermore, the separation between signal and background can be enhanced by the combination of various observables and hence improve the sensitivity to $0\nu\beta\beta$ decay.
The different kinematical variables measured by NEMO-3 are combined into a multi-variate analysis using the Boosted Decision Tree (BDT) implemented in the TMVA package~\cite{2007physics...3039H}.
In addition to the total energy ($E_{Tot}$), the observables considered to enhance signal and background discrimination with BDT are the higher ($E_{Max}$) and lower ($E_{Min}$) electron energies, the asymmetry of the two energies ($E_{Asym}$), the track lengths associated to the higher ($L_{Max}$) and lower ($L_{Min}$) energy electrons, the opening angle between the two tracks ($\cos(\theta)$) and the internal probability distribution ($P_{Int}$).
Dedicated BDTs are trained for each mechanism using 20~\% of the available MC statistics. 
%
%
The remaining 80~\% of the MC statistics are used to test the BDTs against over-training and to evaluate the classifier performance.  
The BDT algorithm returns a score value distributed between $-$1, for background-like events, and $+$1 for signal-like events.
The good agreement between signal selection efficiencies measured on the training and testing samples is a good indication that the BDTs are not over-trained.
As expected, the variable $E_{Tot}$ is the most discriminating variable for all mechanisms, with a BDT rankings of about 20~\%.
Other variables have relatively similar ranking, ranging between 10~\% to 15~\%.
Fig.~\ref{fig:BDT} shows the distribution of the output score of the BDT trained for the search of $0\nu\beta\beta$ decays via the exchange of a light Majorana neutrino. 
The data are compared to the expected MC model normalised to the values obtained in Sec.~\ref{sec:bkg} and \ref{sec:2nu} and summarised in Tab.~\ref{tab:bkg}.
Since no significant excess of data over the expected background is observed, the BDT score distribution is used to set lower limit on the half-life of the decay at 90~\% C.L. using the modified frequentist approach of~\cite{Junk:1999kv}.
The systematic uncertainties considered for the measurement of the $2\nu\beta\beta$ decay half-life are accounted for in the limit setting procedure fluctuating the background and the signal distributions by random scale factors generated according to Gaussian distributions.
The median and $\pm1\sigma$ expected lower limits on half-lives for the different lepton number violating mechanisms 
are shown in Tab.~\ref{tab:exp_limit} for the BDT and for the total electron energy distribution considered alone.
Compatible sensitivities are obtained with both methods for the different mechanisms considered. 
\begin{table}
\caption{Median and $\pm1\sigma$ expected lower limits on half-lives of lepton number violating processes (in units of $10^{23}$~y) at the 90\% C.L. obtained from the BDT and from the total energy distribution.\label{tab:exp_limit}}
\begin{ruledtabular}
\begin{tabular}{c|ccc|ccc}
\multirow{2}{*}{Mechanism}       & \multicolumn{3}{c|}{Expected (BDT)} & \multicolumn{3}{c}{Expected ($E_{Tot}$)} \\ 
  		                         & $-1\sigma$ & Median & $+1\sigma$    & $-1\sigma$ & Median & $+1\sigma$          \\ \hline
$\langle m_{\nu} \rangle$        & $0.7$      & $1.1$  & $1.5$         & $0.8$      & $1.1$  & $1.6$               \\
$\langle \eta \rangle$           & $0.7$      & $1.0$  & $1.4$         & $0.7$      & $1.0$  & $1.4$               \\
$\langle \lambda \rangle$        & $0.5$      & $0.7$  & $1.0$         & $0.4$      & $0.6$  & $0.8$               \\
$\langle g_{\chi^{0}} \rangle $  & $0.07$     & $0.11$ & $0.15$        & $0.07$     & $0.11$ & $0.15$              \\
\end{tabular}
\end{ruledtabular}
\end{table}
\begin{table}
\caption{Limits at the 90\% C.L. on half-lives (in units of $10^{23}$~y) and lepton number violating parameters.\label{tab:obs_limit}}
\begin{ruledtabular}
\begin{tabular}{rcl}
Mechanism                        & Obs. $T_{1/2}^{0\nu}$ limit & Parameter value               \\ \hline
$\langle m_{\nu} \rangle$ & \multirow{2}{*}{$\geq1.0$}  & $\leq(1.4-2.5)$~eV            \\ 
$\lambda^{'}_{111}$              &                             & $\leq0.1\times f$             \\
$\langle \eta \rangle$           & $\geq1.1$                   & $\leq(2.5-11.9)\times10^{-8}$ \\
$\langle \lambda \rangle$        & $\geq0.6$                   & $\leq(3.6-43.0)\times10^{-6}$ \\
$\langle g_{\chi^{0}} \rangle $  & $\geq0.085$                 & $\leq(5.2-9.2)\times10^{-5}$  \\
\end{tabular}
\end{ruledtabular}
\end{table}
The observed 90\% C.L. limits on half-lives and on the lepton number violating parameters are summarised in Tab.~\ref{tab:obs_limit}.
The observed limits are well within the expected $1\sigma$ intervals for all the mechanisms considered.
A lower limit of $1.0\times10^{23}$~y is obtained on the light Majorana neutrino exchange. 
The $0\nu\beta\beta$ half-life limit is converted into an upper limit on the effective Majorana neutrino mass $\langle m_{\nu} \rangle < 1.4-2.5$~eV, using the NMEs from~\cite{PhysRevC.87.045501,PhysRevC.91.024613,PhysRevC.91.034304,PhysRevLett.105.252503,PhysRevC.91.024316}, the Phase Space Factors from~\cite{PhysRevC.85.034316,Pahomi:2014kpa} and $g_{A} = 1.27$.
In the case of the exchange of R-parity violating supersymmetric model, the same limit on the half-life is used to extract an upper limit on the coupling constant $\lambda^{'}_{111}$ assuming the decay proceeds via the exchange of a gluino.
Using the NME from~\cite{Faessler:1997pa} the limit $\lambda^{'}_{111}<0.1\times f$ is obtained, where $f =\left( \frac{m_{\tilde{q}}}{1\mbox{TeV}} \right)^{2} \left( \frac{m_{\tilde{g}}}{1\mbox{TeV}} \right)^{1/2}$  
and $m_{\tilde{q}}$ and $m_{\tilde{g}}$ are the squark and the gluino masses.
In case of the exchange of a right-handed currents the summed energy spectra do not differ significantly from the light neutrino exchange mode. 
The different kinematics of the two electrons considered in the BDT however allows one to enhance the signal to background separation and improve the limits with respect to the total electron energy considered alone.
The resulting limits on the half-lives are $1.1\times10^{23}$~y for the $\langle \eta \rangle$ term and  $0.6\times10^{23}$~y for the $\langle \lambda \rangle$.
Limits on the coupling constants are obtained from the half-life limit using calculations from~\cite{Suhonen:1998ck}  and correspond to $\langle \eta \rangle<(2.5-11.9)\times10^{-8}$ and $\langle \lambda \rangle<(3.6-43.0)\times10^{-6}$.
Finally, in the case of a single majoron emission the obtained limit on the half-life of the decay is $8.5\times10^{21}$~y. 
The limit is more than one order of magnitude lower than other mechanisms due to the signal spectral shape which extends to lower energies dominated by background.
Using the same NMEs adopted for the light Majorana exchange, the phase space factor from~\cite{PhysRevC.91.064310,PhysRevC.92.029903} and $g_{A} = 1.27$, the upper limit on the coupling constant with the majoron is $\langle g_{\chi^{0}}\rangle<(5.2-9.2)\times10^{-5}$.
%
Fig.~\ref{fig:0nubb} shows the distribution of the total electron energy in the two-electron channel above 2~MeV with the expected signal shapes for the different mechanisms normalised to the observed limits.
%
\section{Conclusions \label{sec:conclusion}}
%
The NEMO-3 experiment studied the $\beta\beta$ decay of $^{116}$Cd to the ground state of $^{116}$Sn with  410~g of isotope and a total exposure of 5.26~y.
The half-life of the $2\nu\beta\beta$ decay has been measured to be $ T_{1/2}^{2\nu}=[2.74\pm0.04\mbox{(stat.)}\pm0.18\mbox{(syst.)} ]\times10^{19}~\mbox{ y}$ in the Single State Dominance hypothesis and $T_{1/2}^{2\nu}=[2.96\pm0.04\mbox{(stat.)}\pm0.20\mbox{(syst.)}] \times10^{19}~\mbox{ y}$ in the Higher State Dominance hypothesis.
Thanks to the unique capability of the NEMO-3 detector to fully reconstruct the complete kinematics of the $2\nu\beta\beta$ decay, a dedicated study on the shape of the single electron energy distribution has been performed to discriminate between the SSD and HSD hypothesis.
The NEMO-3 data are in favour of SSD but there is not enough sensitivity to exclude the HSD due to the limited statistics.
The result of this work represents the most precise measurement of the $2\nu\beta\beta$ decay half-life of $^{116}$Cd.
A search for $0\nu\beta\beta$ decay of $^{116}$Cd has been performed in the same dataset.
No significant excess of data over the expected background events is observed.
The lower limit on the half-life for the light Majorana neutrino exchange mechanism has been determined to be $T_{1/2}^{0\nu} > 1.0\times10^{23}$~y at the 90~\% C.L..
The limit on the half-life corresponds to an upper limit on the effective Majorana neutrino mass  $\langle m_{\nu} \rangle < 1.4-2.5$~eV depending on the NME values considered.
Limits have also been set on R-parity violating supersymmetry, right-handed current and majoron emission models.
Due to a lower exposure, these limits are about one order of magnitude weaker than word leading best limits obtained on $^{100}$Mo~\cite{Arnold:2015wpy}, $^{76}$Ge~\cite{PhysRevLett.111.122503}, $^{136}$Xe~\cite{Asakura:2014lma} and $^{130}$Te~\cite{PhysRevC.93.045503}. 
The result of this work is however competitive with the best limits set for $^{116}$Cd~\cite{PhysRevC.68.035501,Danevich:2016eot,Danevich:2000cf}. 
%
\begin{acknowledgments}
We thank the staff of the Modane Underground Laboratory for their technical assistance in running the experiment. 
We acknowledge support by the grant agencies of the Czech Republic, CNRS/IN2P3 in France, RFBR in Russia, APVV in Slovakia, STFC in the U.K. and NSF in the U.S.
This work was supported by the Labex ENIGMASS, Investissements d'Avenir.
\end{acknowledgments}

\bibliographystyle{bibstyles/h-physrev}
\bibliography{biblio.bib}
\end{document}